\documentclass[reprint,
 amsmath,amssymb,aps,prb,
]{revtex4-2}

\usepackage{graphicx}
\usepackage{dcolumn}
\usepackage{bm}
\usepackage{hyperref}

\usepackage{amssymb,amsthm}
\usepackage{braket}
\usepackage{dsfont}
\usepackage{xcolor}
\renewcommand{\atop}[2]{\genfrac{}{}{0pt}{}{#1}{#2}}

\newcommand{\abs}[1]{\left\vert#1\right\vert}

\renewcommand{\i}{\mathfrak{i}}

\begin{document}
\title[Hartree Method for Molecular Polaritons]{Hartree Method for Molecular Polaritons.}
\author{Vladimir Al. Osipov}
\email{Vladimir.Al.Osipov@gmail.com}
\affiliation{H.I.T.-Holon Institute of Technology, 52 Golomb Street, POB 305
Holon 5810201, Israel}

\author{Boris Fainberg}\email{Fainberg@hit.ac.il}
\affiliation{School of Chemistry, Tel Aviv University, Tel Aviv 69978, Israel}
\affiliation{H.I.T.-Holon Institute of Technology, 52 Golomb Street, POB 305
Holon 5810201, Israel}

\date{\today }

\begin{abstract}
The formation of the composite photonic-excitonic particle, known as a polariton, is a phenomenon emerging in materials possessing strong coupling to light. The organic-based materials besides the strong light-matter coupling also demonstrate strong interaction of electronic and vibrational degrees of freedom. We study the vibration-assisted polariton wavefunction evolution treating both types of interactions as equally strong. Using the multiconfiguration Hartree approach we derive the equations of motion for the polariton wavefunction, where the vibration degrees of freedom interact with the polariton quantum field through the mean-field Hartree term. For the conventional quadratic polariton Hamiltonian and the Holstein-like vibration Hamiltonian (Tavis–Cummings–Holstein model), the obtained equations are in one-to-one correspondence with the original Schr\"odinger equation. In the second part of the article, we show that our theory reproduces the physical properties of the polariton light emission spectrum. In particular, the theory explains experimental observations of the molecular Stokes shift in the polariton fluorescence spectra in the systems with strong light-matter coupling. We also investigate the behaviour of the polariton wavefunction in the vicinity of the anticrossing point and demonstrate that the Hartree term can produce an infinite potential barrier of a dynamical origin, which is responsible for the formation of the mixed upper-lower polariton states. The nonlinear nature of the polariton theory reflects their collective behaviour. We expect that the multiconfiguration Hartree approach being applied to polaritons and similar systems will result in a manifestation of new physical phenomena.
\end{abstract}

\maketitle

\section{Introduction}

When dye molecules form a microstructure such as a nanofiber crystal \cite{Takazawa10}, or a molecular solution placed into a microcavity, the strong light-matter interaction can lead to the formation of composite photon-exciton particles, known as exciton-polaritons, see~\cite{Ebbesen2016,HWMB2019,KavokinBook,Li18324} and references therein. The systems with strong coupling to light draw a lot of attention due to their potential to cause changes in chemical reaction rates~\cite{HSGDE2012}, to form high-temperature polariton Bose-Einstein condensate~\cite{KavokinBook,SFW1998,KRK2006,DCTH2006}, and to show long-range particle propagation~\cite{Takazawa10, RAGS2018, HKDQNMF2020}. The concept, generally used to describe the polaritons in the organic-based devices, deals with the Frenkel excitons~\cite{Agranovich09}. This type of excitons possesses the property to be localized on a few molecules (the typical size $\sim $10\AA ), which, in particular, defines their stability (the binding energy $\sim 1$eV). A stronger coupling is achieved due to the large oscillator strength of the dye molecules. Entanglement of the excitonic states with certain light modes formed in the microcavity or with the free radiation modes splits the linearly growing with respect to the wavevector~$q$ light dispersion curve in the point of its crossing (the anticrossing point, AP) with almost constant exciton energy~$\omega_{ex}$. The gap width  between the lower and the upper polariton dispersion branches (Rabi splitting) is governed by the strength of the light-matter interaction parameter $g$. In organic materials, $g$ can reach significant values, up to $1$eV~\cite{Takazawa10,HWMB2019}. The above picture becomes more complex when one accounts for the interaction of the molecular electrons with the vibrational degrees of freedom, the inherent property of organic materials. The vibrations are known to assist the exciton-polariton stability and participate in such dynamic rearrangements of the polariton systems as bose-like condensation~\cite{KavokinBook}, transport~\cite{Takazawa10, HKDQNMF2020,Fainberg17APL}, and relaxation~\cite{CMCTAKL2011,HPTTGT2021,FMSWMBPZ2021,BHGHT2017,LRA2004,CMCTAKL2011}.
Therefore, when describing complex processes involving polaritons, both the interaction of light with molecules and the interaction of electron density with vibrations of the molecular cores must be treated on equal footing~\cite{Toyozawa59, RSG2019,SN2019,WFG2016,Fainberg18Advances,Fainberg19JPCC}.

Vibrational degrees of freedom in crystals form the number of acoustic and optical phonon modes, which accounting is essential for describing semiconductor-based polaritonic devices. In the solutions of the organic dye molecules, the vibrations in the vicinity of each optically active centre can be considered as if they were independent, and the other vibration degrees of freedom in the material or the solution effectively form a thermal bath. On the other hand, the large displacement of the equilibrium nuclear positions of the low-frequency optically active vibrations under the optical electronic transition leads to their excitation with large quantum numbers. In the monomolecular spectra, the low-frequency vibration modes reveal themselves in the Stokes shift between the emission and absorption peaks. The high-frequency optically active vibrations become visible in the form of the vibrational progression. Such strong effects must also influence the polariton spectra. The influence of the high-frequency vibration modes on polaritons appeared to be more diverse. Its study led to some new interesting phenomena. The main peaks of the polariton fluorescent spectra are naturally associated with the energies of the upper and lower dispersions. The presence of the additional peaks and thermal broadening in the polariton spectra was studied in several works~\cite{BHGHT2017, HF2004, Rocca09, HS2017}. Recently it was shown that the electron-vibrational interaction could result in the formation of the non-Markovian Fano resonances and the motional narrowing of the exciton-polariton luminescence spectrum~\cite{Fainberg22JPCA}. In a number of articles, it was also shown that the molecular Stokes shift is one of the system parameters which can influence the polariton dynamics~\cite{HPTTGT2021,TBKPN2017,LBVAS1999}. 

The approaches used to describe the polariton-vibration system were mainly based on phenomenological arguments. The aim of the present research is in developing a rigorous, derived from first principles, approach for the description of the polariton-vibration system evolution. In particular, we study the vibration-assisted quantum evolution of the single-polariton wavefunction. It is worthy to note that for the solution of the transport problem, accounting of the polariton-polariton interaction~\cite{GOPMKD2009}, description of other dynamical processes~\cite{WFG2016}, and also for developing of the multidimensional spectroscopy methods~\cite{XRDWLSOYX2018,FMSWMBPZ2021}, the knowledge of
the polariton wavefunction evolution and its spatial propagation is of
importance. A number of attempts to describe the polariton wavefunction have been done without accounting of vibrations~\cite{S1981, QAB1986}.
In the present article, starting from the Tavis–Cummings–Holstein model Hamiltonian~\cite{CRLK2014} we derive a set of equations for the polariton-vibration wavefunction evolution in the Hartree approximation. Using these equations we consider several problems
admitting approximate analytic solutions.

The vibronic coupling in a molecule interrelates the electronic and nuclear vibrational motion. In theoretical chemistry, the vibronic coupling is often neglected within the Born-Oppenheimer approximation. The couplings become crucial to the understanding of nonadiabatic processes, especially near AP, when the energy gap order of magnitude is comparable with the oscillation quantum energy. The large magnitude of the vibronic coupling near AP allows the wave function to propagate from one adiabatic potential energy surface to another, giving rise to a nonadiabatic phenomenon such as radiationless decay in molecular systems. The vibronic coupling can also form a singularity of the conical intersection type. This type of singularity is responsible for appearance of a non-zero geometric phase, which, in the context of molecular dynamics, was discovered by Longuet-Higgins~\cite{HOPS1958}. In this case it becomes essential to account for the quantum interference of the system wavefunction with itself. In the context of the polariton-type systems this effect was discussed in~\cite{GM2020}.

In our work, to describe the vibration degrees of freedom we use the language of coherent states.  With a few relatively simple exceptions~\cite{TSM1998}, the direct calculation of the quantum transitions assisted by vibrations within the coherent state framework~\cite{BBGK1971} is not common due to the difficulties associated with their evaluation. To this end, several theoretical approaches have been developed, including the method of coupled coherent states~\cite{SC2004}. The concept implies that the quantum trajectories are allowed to explore the phase-space wider than the zero-vibration space. Having been equipped by a phase, the quantum
trajectories start to interfere with each other. In this sense, the quantum effects can be thought of as arising from the interaction of the
trajectories. This method belongs to a wider class of methods, which solve
the Schr\"{o}dinger equation in a time-dependent basis set and the
time-evolutions of both the basis vectors and that of the wavefunction
expansion coefficients are determined from the Dirac-Frenkel (sometimes Dirac-Frenkel/McLachlan) variational principle.

The coupled coherent states method is mainly used to describe the quantum evolution of a single molecule in an external field or without it. Contrary to the single-molecule models, the polariton quasi-particle is a composite particle and its quantum wave packet is spread over the whole microscopic sample (the polariton wavelength at AP is about 50nm) and thus includes the quantum states of all molecules in the sample. The general nonlinear equations of polariton motion,  obtained in the next section, include the forces acting from the side of each molecule on the polariton particle and also the backward action of the polariton on the molecular vibrations. Such a general model has to describe the whole variety of physical effects taking place in the system, including the effects of decoherence. The influence of different effects can be singled out by the special choice of the molecular Hamiltonian and, which is not less important, by an appropriate choice of the method for solving the nonlinear equation. In the present article, we use the multiconfiguration Hartree approach to formulate the equations of polariton motion in the mean-field approximation. In this approach the set of equations of motion split into two parts. The first part are the classical Newton equations, which describe the evolution of the vibration degrees of freedom. The presence of polaritons in this equation is taken into account by the Hartree term, which enters as a classical force resulting from the quantum averaging of the polariton field. The second part of the equations has the form of the Schr\"{o}dinger equation written for the coefficients of the polariton wavefunction expansion (they correspond to Hopfield coefficients in the standard polariton theory). These equations also contain a term proportional to the Hartree amplitude~\cite{SOBook}, which makes the equations essentially nonlinear. It turned out that the reformulation of the initial Schr\"{o}dinger equation into the set of equations of the vibration-assisted polariton motion is exact in the case of the canonical quadratic polariton Hamiltonian and the Holstein-like vibration Hamiltonian~\cite{HOLSTEIN1959325}. The nonlinearity of the polariton equations reflects their collective behaviour. We expect that the nonlinear effects can result in a manifestation of new physical phenomena in the polariton systems and also in similar multimolecular systems. 

The article is structured in the following way: In the next section (section~\ref{EqMotSec}), we derive equations of motion for the vibration-assisted polariton wavefunction. To do this, first (section~\ref{sec1}), we introduce the basis set of quantum polariton states for the quadratic polariton Hamiltonian without vibrations.  To describe the vibrations in section~\ref{sec2} we introduce the basis of time-dependent coherent states. Varying the Schr\"odinger equation formulated for a model polariton-vibration Hamiltonian we derive
the semiclassical equations of polariton evolution (section~\ref{sec3}). 
In the section~\ref{Sec32} we solve the system of polariton equations of motion in the quasi-diagonal approximation to calculate the polariton fluorescent spectrum, some details of calculation are given in the appendices~\ref{AppB} and~\ref{AppC}. The equations of motion in the vicinity of AP and also dynamical formation of a potential barrier separating the mixed states from the pure polariton states are discussed in the section~\ref{sec21}. The results of our research are discussed in the section~\ref{Sec3}.

\section{Equations of motion for the vibration assisted polariton
wavefunction}
\label{EqMotSec}

\subsection{The basis set of the polariton Hamiltonian}

\label{sec1}  In the present work, we focus on the type of systems in which in the first approximation the dipole-dipole interactions between molecules can be neglected. An example of such a system is the solution of enhanced green fluorescent protein (eGFP). The actual fluorophore of FPs is enclosed by a nano-cylinder that consists of eleven $\beta $-sheets \cite{Gather_Yun14,Gather16}. This protective shell acts as a natural ``bumper'' and prevents close contact between fluorophores of neighbouring FPs, limiting the intermolecular energy migration even at the highest possible concentration. The intermolecular interactions impart~\cite{HS2018} the momentum $q$-dependence to the exciton dispersion $\omega _{ex}$. In the case of weakly coupled molecules the short-range Frenkel exciton effective mass is large and we can neglect the $q$-dependence in $\omega _{ex}$. Note that there are examples of systems, where the dipole-dipole interaction (excitonic coupling) is suppressed even in the crystal phase~\cite{Gather16,Silbey70,OCKGTR1996}. To this end, we also add that below we consider the polariton operators as those that satisfy the Bose statistics, which is an assumption taking place at a low density of excitations. Discussion regarding the admissibility of such approximation one can find in Refs.~\cite{CP2008,CP2009}. It was shown there that the composite nature of the Frenkel excitons is responsible for the excitation transfer. In what follows we consider a single polariton wavefunction when the non-bosonic corrections nullify.

\begin{figure}
\centering
\begin{tabular}{lc}
a) & \includegraphics[scale=0.25]{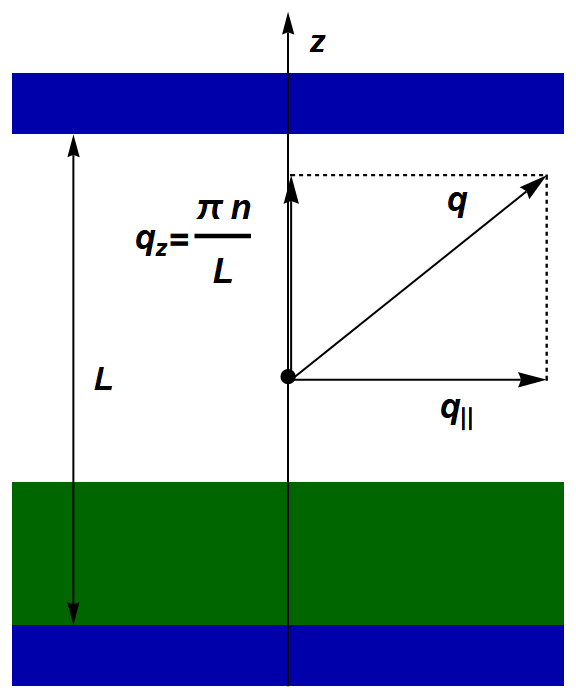}\\
b) & \includegraphics[scale=0.35]{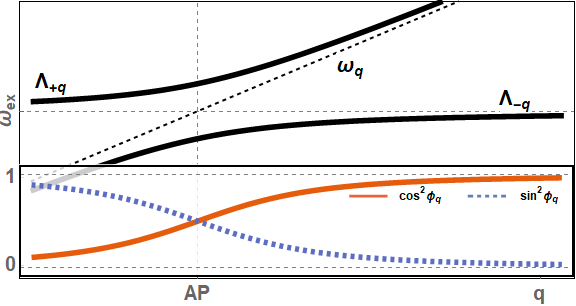}
\end{tabular}
\caption{\small a) The molecular substance (green area) containing a
thin layer of the molecular solution is placed between two cavity mirrors
(blue rectangles) with the distance $L$ between them. The light modes formed in the microcavity have the wavevector $q$ with the in-plane,  $q_{||}$,  and transverse,  $q_{z}$, components, $q=(q_{||},q_{z})$. The transverse component $q_{z}$ can have only discrete
values marked by $n=1,2,\dots $; b) The polariton dispersion curves $\Lambda _{\pm q}
$ and the Hopfield coefficients $\cos ^{2}\protect\phi _{q}$, $\sin ^{2}%
\protect\phi _{q}$ (in the insert) plotted v.s. the wavevector $q$. AP
denotes the anticrossing point $\protect\omega _{ex}=\protect\omega _{q}$.}
\label{fig2a}
\end{figure}

In this section, we introduce the polariton basis vector set for the basic model of the polariton Hamiltonian, $\hat{H}_{pol}$. For systems without any distinguished spatial directions when the light scattering from the inhomogeneities of the medium can be neglected, one can work with the basic model polariton Hamiltonian, $\hat{H}_{pol}$, which diagonalized form is quadratic in the upper, $Q_{q}$, and the lower, $P_{q}$, polariton operators~\cite{Agranovich03},
\begin{equation}
\hat{H}_{pol}=\hbar \sum_{q}\Lambda _{+q}Q_{q}^{\dag }Q_{q}+\Lambda
_{-q}P_{q}^{\dag }P_{q},  \label{HamP}
\end{equation}%
The polariton operators and the polariton energies, $\Lambda _{\pm q}$, depend on the wavevector $q$.  The polariton dispersion relations are known to be the solution of a quadratic equation and expressed in terms of the exciton energy, $\omega _{ex}$, and the photon energy, $\omega _{q}$, 
\begin{equation}
\Lambda _{\pm q}=\frac{1}{2}\left( \omega _{q}+\omega _{ex}\pm \sqrt{(\omega_{q}-\omega _{ex})^{2}+4g^{2}}\right),  \label{Lambd}
\end{equation}
The Rabi splitting, i.e. the width of the gap between the upper and the
lower polariton branches is governed by the light-matter
interaction strength constant $g$. The polariton operators $Q_{q}$ and $P_{q}$ are expressed in terms of the material operators by means of the unitary transformation parametrized by the ``Hopfield
angle'' $\phi _{q}$ (see fig.~\ref{fig2a} b), 
\begin{eqnarray}
Q_{q} &=&\cos \phi _{q}A_{q}-\i \sin \phi _{q}\frac{1}{\sqrt{N}}%
\sum_{m}e^{-\i qm}b_{m};  \label{Pop} \\
P_{q} &=&\sin \phi _{q}A_{q}+\i \cos \phi _{q}\frac{1}{\sqrt{N}}%
\sum_{m}e^{-\i qm}b_{m},  \label{Qop}
\end{eqnarray}
where $N$ is the total number of molecules and  $q m$ denotes the scalar product of the wavevector $q$ and the radius-vector pointing at the optical transition center of the $m$-th molecule. The operators $A_{q}$ in eqs.~(\ref{Pop}),~(\ref{Qop}) are the boson annihilation operators of a photon in the mode $q$. The operators $b_{m}$ are the annihilation operators of the excited state at the $m$-th molecule.  The exciton annihilation and creation operators are known to be paulions (or composite bosons, according to the terminology used in Refs.~\cite{CP2008,CP2009}): they posses the fermion properties, $[b_m,b_m^\dag]_+=1$, at one and the same site, and commute for different sites, $[b_m,b_{m'}^\dag] = 0$ when
$m\ne m'$. Under the assumption of a small density of the excitations, the operators $b_m$ approximately (up to the order 1/N) satisfy the bosonic commutation relations $[b_m,b_{m'}^\dag] = \delta_{m,m'}$, and so do the polariton operators  $[Q_{q},Q_{q'}^{\dag}]=[P_{q},P_{q'}^{\dag }]=\delta _{q,q'}$. The substitution of paulions by bosons for small excitation densities historically can be traced back to the method of approximate second quantization developed in the theory of magnetism~\cite{Tyablikov67}. The paulion state can be either occupied or unoccupied, whereas the occupation numbers for bosons can be any positive integer number. Therefore, the replacement of paulions by bosons fails when the number of bosons is larger than $1$. In the case of nonlinear optical effects this is avoided by adding into the quadratic ``boson'' Hamiltonian (eq.~\ref{HamP})  the operator of kinematic interaction, which includes the terms of the fourth and higher orders~\cite{Agranovich09,Litinskaya08,Zoubi14}. The operator of kinematic interaction results in a nonlinear interaction between the bosons. In this relation, a useful and straightforward method for accounting of the multi-Frenkel exciton states can be found in Ref.~\cite{CP2008,CP2009}.  In our theory, where the interaction between polaritons is caused by vibrations, the nonlinear equations appear even for the single-polariton state describing the single-exciton processes. The solution of nonlinear equations is a non-trivial problem by itself, so to focus on a new physics related to our nonlinear theory we exclude the additional nonlinearities generated by the multiexciton states in the nonlinear optical processes. Thus we restrict the application of our theory to the linear optical processes for which it is sufficient to take into account the single-exciton states.

A few remarks have to be done at this point: 

I. The Hopfield angle $\phi _{q}$ in eqs.~(\ref{Pop}),~(\ref{Qop}) is defined through the relation $\cos 2\phi _{q}=\frac{\omega _{q}-\omega _{ex}}{\Lambda _{+q}-\Lambda _{-q}}$. Its value ranges from $0$ at large $q$, $q\rightarrow \infty $, to some value close but smaller than $\pi /2$ at $q=0$. The wavevector satisfying the AP position, $\omega_q=\omega_{ex}$, corresponds to $\phi_{q}=\pi /4$.  

II. The problem implies the symmetry with respect to the generic change of the momentum sign, so that all the equations have to be invariant under the transformation $q\rightarrow -q$, and, in particular, $\omega_{q}=\omega_{-q}$ and $\phi_{q}=\phi_{-q}$. 

III. In our formulation we use the rescaling, where $m$ is an integer-valued vector and the dimension units are absorbed by the wavevector $q$ and also by the energy units. In the free space the photons dispersion $\omega_q$ is linear in the wavevector absolute value. In our notations the free space photon energy has the form $\omega_q=c\abs{q}/n_0<\ell >$, where $c/n_0$ is the speed of light in the medium and $<\ell >$ is the mean distance between the molecules. Below, for convenience, we omit $<\ell >$, bearing in mind that $<\ell >$ is cancelled in the final formulas, and $q$ is measured in conventional units. In the case when the active media is placed into a microcavity the wavevector values are bounded from below by the wavevector $q_{z}$ of the eigenmode excited in the resonator and $\omega_q=\frac{c}{n_0}\sqrt{q_{z}^{2}+q_{||}^{2}}$ (see fig.~\ref{fig2a}).

IV. The single-polariton states of the polariton Hamiltonian $\hat{H}_{pol}$ (eq.~\ref{HamP}) is composed of the vectors which we denote by  $\ket{q,u}$. They are distinguished by the parameters $q$, $u$. Each of $\ket{q,u}$ describes an upper ($u=1$) or lower ($u=0$) polariton excited with the momentum $q$,
\begin{equation}\label{basisPQ}
\ket{q,0}=P_q^\dag\ket{0},\quad \ket{q,1}=Q_q^\dag\ket{0}. 
\end{equation}
The vector $\ket{0}$ denote the ground state of the system with zero polaritons. The non-zero matrix elements of the polariton Hamiltonian are 
\begin{equation}
\bra{q,u}\hat{H}_{pol}\ket{q,u}=(1-u)\Lambda _{-q}+u\Lambda _{+q}.
\label{polHamEigS}
\end{equation}

V. Accounting of the vibrations  shifts the energy of the molecular optical transition by half of the Stokes shift, $\omega _{st}/2$. In the next section (section~\ref{sec2}) it is implied that $\omega _{ex}\rightarrow \omega_{ex}+\omega _{st}/2$ in eq.~(\ref{Lambd}) and in the related equations. 

\subsection{The vibration Hamiltonian and the extended basis set of the
polariton wavefunctions}

\label{sec2} The total Hamiltonian of the molecular system with polaritons, $%
\hat{\mathcal{H}}$, contains two contributions: the polariton part, $\hat{H}%
_{pol}$, which was discussed in the previous section (eq.~\ref{HamP}), and
the vibrational part, $\hat{H}_{vib}$, which describes interaction of
electrons with vibrations: 
\begin{equation}
\hat{\mathcal{H}}=\hat{H}_{pol}+\hat{H}_{vib}.  \label{Hamtot}
\end{equation}%
The vibrational part of the Hamiltonian, is modelled by the standard
electron-vibration Holstein-like~\cite{HOLSTEIN1959325} Hamiltonian,
\begin{multline}\label{Hvib}
\hat{H}_{vib}=\sum_{m,\mu }\bigg[\hat{H}_{bath}\big(c_{m,\mu }^{\dag },c_{m,\mu }\big)+  \hbar \Omega _{\mu }c_{m,\mu }^{\dag
}c_{m,\mu }\\-\hbar \Omega _{\mu }X_{\mu }(c_{m,\mu }^{\dag }+c_{m,\mu
})b_{m}^{\dag }b_{m}
\bigg] .  
\end{multline}
It is assumed that the electron transition in the $m$-th molecule is coupled to a number of quantum harmonic oscillations of the molecular backbone with various energy quanta $\hbar \Omega _{\mu }$ indexed by $\mu =0,1,\dots $. The frequencies $\Omega _{\mu }$ and the oscillator equilibrium coordinates shifts $X_{\mu }$ in the excited electronic state are equal for all molecules. The optically active vibration modes (boson operators $c_{m,\mu }$ and $c_{m,\mu }^{\dag }$, $[c_{m,\mu },c_{m',\mu'}^{\dag }]=\delta_{m,m'}\delta_{\mu,\mu'}$) interact with the dark modes via the thermal bath Hamiltonian $\hat{H}_{bath}$. We assume that this term also includes, if necessary, the interactions of vibrations localized at different molecules. 
Remind here that the excitation energy has to be shifted by the half of the Stokes shift $\omega _{st}\equiv 2\sum_{\mu }\Omega _{\mu
}X_{\mu }^{2}$, i.e. we imply that $\omega _{ex}\rightarrow \omega_{ex}+\omega _{st}/2$ in eq.~(\ref{Lambd}) and in the related equations.

The basis of coherent states provides a convenient description for vibration degrees of freedom~\cite{BBGK1971}. Each coherent state, $\ket{\sigma}$, is parametrized by multidimensional complex-valued vector $\sigma$. It encodes the coherent state center, i.e. the classical coordinate $x$ and the classical momentum $p$, namely $\sigma=x+\i p$. The vibration operators act on the basis vectors as follows: $c\ket{\sigma}=\sigma \ket{\sigma }$, $\bra{\sigma }c ^\dag=\bra{\sigma }\sigma^*$ and the normalized coherent state have the representation 
\begin{equation}  \label{cohState}
\ket{\sigma }=e^{-\frac{1}{2}\left\vert \sigma \right\vert^2}e^{\sigma
c^\dag }\ket{0 },
\end{equation}
where $\ket{0 }$ is the ground state of the corresponding oscillator.

As we mentioned in the previous section, we consider only the single-polariton state, which is described by two types of vectors: $\ket{q,0}$, and $\ket{q,1}$. Thus, our working basis consists of the direct
products of the polariton and vibrational states ($u=0,1$) 
\begin{equation}
\ket{\bm \sigma,q,u}=\ket{\bm \sigma}\ket{q,u},\quad \ket{\bm \sigma}%
=\bigotimes_{m,\mu }\ket{\sigma_{m,\mu}}.  \label{ketsigmaa}
\end{equation}%
The scalar product of the vectors defined above is 
\begin{multline}
\braket{\bm \sigma',q',u'|\bm\sigma,q,u}=\braket{\bm\sigma'|\bm\sigma}%
\braket{q',u'|q,u}\\
=\exp \left( -\frac{1}{2}\sum_{m,\mu }\left\vert \sigma
_{m,\mu }^{\prime }-\sigma _{m,\mu }\right\vert ^{2}\right) \delta
_{q^{\prime },q}\delta _{u^{\prime },u}.
\end{multline}
The completeness and orthogonality relation for the extended basis (eq.~\ref{ketsigmaa}) takes the form 
\begin{equation}
\mathds{1}=\sum_{\bm\sigma }\sum_{q}\sum_{u=0,1}\ket{\bm\sigma,q,u}%
\bra{\bm\sigma,q,u}.
\end{equation}
Here and below the sum $\sum_{\bm\sigma }$ denotes the integration $\int
d^{2}\sigma _{m,\mu }/\pi $ over each component of the vector $\bm\sigma $.

\subsection{Polariton wavefunction evolution in the time-dependent basis and equations of motion}

\label{sec3} The basis of vectors (eq.~\ref{ketsigmaa}) defined in the previous section does
not diagonalize the total Hamiltonian $\hat{\mathcal{H}}$.
To describe the evolution of the wavefunction we make use an Anzatz, which
states that the basis of the coherent states is time-dependent, i.e. in
addition to the time-dependence of the expansion coefficients $\mathcal{C}(\bm \sigma,q,u|t)$, we assume that the basis vectors also depend on time, $\ket{\bm \sigma(t),q,u}$. The wavefunction of the system can be standardly expanded in this basis 
\begin{equation}  \label{wavefunction1}
\ket{\Psi(t)}=\sum_{\bm\sigma,q,u}\mathcal{C}(\bm \sigma,q,u|t)\ket{\bm
\sigma(t),q,u}.
\end{equation}
To work with such wavefunctions we use the approach, which is based on the
Dirac-Frenkel variation principle. This approach is known to be useful for
the description of quantum dynamics in systems with a large number of
vibration degrees of freedom~\cite{SC2004,AR2017,Miller2002,WG2019}. In our
case it allows us to separate the time-evolution of the vibration subsystem
and the quantum evolution of the polariton wavefunction. The wavefunction
(eq.~\ref{wavefunction1}) variation is
{\small \begin{multline}
\bra{\delta\Psi(t)}=\sum_{\sigma,q,u}\bra{\bm\sigma,q,u}\Bigg\{\delta%
\mathcal{C}^*(\bm \sigma,q,u|t) +\mathcal{C}^*(\bm \sigma,q,u|t)\\ \times
\sum_{m,\mu}\left(\delta\sigma^*_{m,\mu}
c_{m,\mu}-\frac{1}{2}\left[\sigma_{m,\mu}
\delta\sigma_{m,\mu}^*+\sigma_{m,\mu}^*\delta\sigma_{m,\mu}\right]\right)%
\Bigg\}.  \label{WFVar}
\end{multline}} 
The second term in eq.~(\ref{WFVar}) results from the variation of the coherent state written in
the form eq.~(\ref{cohState}). Variation of the Schr\"odinger equation $\bra{\Psi}\i\hbar\frac{d}{dt} -\hat{\mathcal{H}}\ket{\Psi}=0$ with respect
to the bra-vector and equating to zero each term proportional to the
independent variations $\delta \mathcal{C}(\bm \sigma',q',u'|t)$, and $\delta\sigma_{m,\mu}$
yields the system of coupled equations, 
{\small \begin{eqnarray}
&&  \bra{\bm\sigma',q',u'}\i\hbar\frac{d}{dt}-\Hat{\mathcal{H}}\ket{\Psi}=0;
\label{varH1} \\
&& \mathcal{C}^*(\bm \sigma',q',u'|t)\bra{\bm\sigma',q',u'}c_{m,\mu}\left(\i\hbar\frac{d}{dt}-\Hat{\mathcal{H}}
\right)\ket{\Psi}=0.  \label{varH2}
\end{eqnarray}}
From eq.~(\ref{WFVar}) we also derive the expression for the time-derivative of
the wavefunction, 
{\small \begin{multline}
\frac{d}{dt}\ket{\Psi(t)}=\sum_{\sigma,q,u}\Bigg\{\dot{\mathcal{C}}(\bm \sigma,q,u|t) + \mathcal{C}(\bm \sigma,q,u|t)\sum_{m,\mu}   
\bigg(\dot{\sigma}_{m,\mu}
c^\dag_{m,\mu}\\-\frac{1}{2}\left[\sigma_{m,\mu} \dot{\sigma}
_{m,\mu}^*+\sigma_{m,\mu}^*\dot{\sigma}_{m,\mu}\right]\bigg)\Bigg\} \ket{\bm\sigma,q,u}.  \label{psivar}
\end{multline}}
The dot symbol, as usually, denotes the time-derivative. Substitution of the above expression~(eq.~\ref{psivar}) into eqs.~(\ref{varH1}),~(\ref{varH2}) leads to two equations which extended forms are 
\begin{widetext}
\begin{multline}  \label{eq1}
\sum_{\bm \sigma}\braket{\bm \sigma'|\bm \sigma}\dot{\mathcal{C}}(\bm %
\sigma,q^{\prime },u^{\prime }|t) =-\frac{\i}{\hbar}\sum_{\bm \sigma,q,u}%
\bra{\bm \sigma',q',u'}\Hat{\mathcal{H}}\ket{\bm \sigma,q,u}\mathcal{C}(\bm %
\sigma,q,u|t) \\
+\frac{1}{2}\sum_{\bm \sigma}\braket{\bm \sigma'|\bm \sigma}\mathcal{C}(\bm %
\sigma,q^{\prime },u^{\prime }|t)\sum_{m,\mu}\left[\sigma_{m,\mu} \dot{\sigma%
}_{m,\mu}^*-\sigma_{m,\mu}^*\dot{\sigma}_{m,\mu}\right] 
-\sum_{\bm\sigma}\braket{\bm \sigma'|\bm \sigma}\mathcal{C}(\bm %
\sigma,q^{\prime },u^{\prime }|t)\sum_{m,\mu}({\sigma^{\prime }}%
^*_{m,\mu}-\sigma_{m,\mu}^*)\dot{\sigma}_{m,\mu};
\end{multline}
\begin{multline}  \label{eq2}
\mathcal{C}^*(\bm \sigma^{\prime },q^{\prime },u^{\prime }|t)\sum_{\sigma}%
\braket{\bm \sigma'|\bm \sigma}\Bigg[\sigma_{m,\mu} \dot{\mathcal{C}}(\bm %
\sigma,q^{\prime },u^{\prime }|t) +\dot{\sigma}_{m,\mu}\mathcal{C}(\bm %
\sigma,q^{\prime },u^{\prime }|t)\Bigg] 
=-\frac{\i}{\hbar}\mathcal{C}^*(\bm \sigma^{\prime },q^{\prime },u^{\prime
}|t)\sum_{\bm \sigma,q,u}\bra{\bm \sigma',q',u'}c_{m,\mu}\Hat{\mathcal{H}}%
\ket{\bm \sigma,q,u}\\ \times\mathcal{C}(\bm \sigma,q,u|t) 
+\frac{1}{2}\mathcal{C}^*(\bm \sigma^{\prime },q^{\prime },u^{\prime
}|t)\sum_{\bm\sigma}\braket{\bm \sigma'|\bm \sigma}\sigma_{m,\mu}\mathcal{C}(%
\bm \sigma,q^{\prime },u^{\prime }|t) \sum_{n,\nu}\bigg(\left[\sigma_{n,\nu} \dot{%
\sigma}_{n,\nu}^*-\sigma_{n,\nu}^*\dot{\sigma}_{n,\nu}\right]-2({\sigma^{\prime }}%
^*_{n,\nu}-\sigma_{n,\nu}^*)\dot{\sigma}_{n,\nu}\bigg) 
\end{multline}
\end{widetext}

Now we have to specify the terms including the Hamiltonian $\hat{\mathcal{H}}$. The system Hamiltonian $\Hat{\mathcal{H}}$ consists of two parts $\hat{H}_{pol}$, and $\hat{H}_{vib}$. The polariton Hamiltonian (eq.~\ref{HamP}) is diagonalized in the basis~(\ref{ketsigmaa}), its non-zero entries are 
{\small \begin{eqnarray}
\bra{\bm\sigma',q',u'}\hat{H}_{pol}\ket{\bm\sigma,q,u}&=&\hbar%
\braket{\bm\sigma'|\bm\sigma}\delta_{q^{\prime },q}\delta_{u^{\prime
},u}\nonumber \\&&\left((1-u)\Lambda_{-q}+u\Lambda_{+q}\right);  \label{A9} \\
\bra{\bm\sigma',q',u'}c_{m,\mu}\hat{H}_{pol}\ket{\bm\sigma,q,u}&=&\sigma_{m,\mu} \hbar 
\braket{\bm\sigma'|\bm\sigma}\delta_{q^{\prime },q}\delta_{u^{\prime
},u}\nonumber \\&&\left((1-u)\Lambda_{-q}+u\Lambda_{+q}\right).  \label{A91}
\end{eqnarray}}

To continue our calculations we make use the general observation, that the
matrix elements $\bra{\bm\sigma',q',u'}\Hat{\mathcal{H}}\ket{\bm\sigma,q,u}$ of a generic normally ordered in $c_{m,\mu}^\dag$ and $c_{m,\mu}$ Hamiltonian $\Hat{\mathcal{H}}$ can be obtained by replacing of $%
c_{m,\mu}^\dag$ and $c_{m,\mu}$ by ${\sigma_{m,\mu}^{\prime }}^*$ and $%
\sigma_{m,\mu}$, respectively, i.e. 
{\small \begin{equation}  \label{HamPol}
\bra{\bm\sigma',q',u'}\Hat{\mathcal{H}}\ket{\bm\sigma,q,u}=%
\braket{\bm\sigma'|\bm\sigma}\bra{q',u'}\Hat{\mathcal{H}}(\bm{\sigma'}^*,\bm%
\sigma)\ket{q,u}
\end{equation}}
and correspondingly 
\begin{multline}  \label{cmHam}
\bra{\bm\sigma',q',u'}c_{m,\mu}\Hat{\mathcal{H}}\ket{\bm\sigma,q,u}%
-\sigma_{m,\mu}\bra{\bm\sigma',q',u'}\Hat{\mathcal{H}}\ket{\bm\sigma,q,u} \\
=\braket{\bm\sigma'|\bm\sigma}\frac{\partial}{\partial{\sigma_{m,\mu}^{%
\prime }}^*}\bra{q',u'}\Hat{\mathcal{H}}(\bm{\sigma'}^*,\bm\sigma)\ket{q,u}.
\end{multline}

The matrix elements of the vibration Hamiltonian can be calculated with the
help of eqs.~(\ref{Pop}),~(\ref{Qop}). The electron operators expressed in
terms of the polariton operators yield 
\begin{multline}
b_m^\dag b_m=\frac{1}{N}\sum_{q^{\prime },q}e^{-\i(q^{\prime }-q)m}(\sin
\phi_{q^{\prime }} Q_{q^{\prime }}^\dag - \cos\phi_{q^{\prime }}
P_{q^{\prime }}^\dag)\\\times (\sin \phi_{q} Q_{q} - \cos\phi_{q} P_{q}).
\end{multline}
The matrix elements of the polariton operator products are
calculated from eq.~(\ref{basisPQ}), namely 
\begin{eqnarray}
\bra{q',u'}Q_{s^{\prime }}^\dag Q_s\ket{q,u}&=&u\,\delta_{u^{\prime
},u}\delta_{s^{\prime },q^{\prime }}\delta_{s,q}; \\
\bra{q',u'}P_{s^{\prime }}^\dag P_s\ket{q,u}&=&(1-u)\,\delta_{u^{\prime
},u}\delta_{s^{\prime },q^{\prime }}\delta_{s,q}; \\
\bra{q',u'}P_{s^{\prime }}^\dag Q_s\ket{q,u}&=&u(1-u^{\prime
})\,\delta_{s^{\prime },q^{\prime }}\delta_{s,q}; \\
\bra{q',u'}Q_{s^{\prime }}^\dag P_s\ket{q,u}&=&u^{\prime
}(1-u)\,\delta_{s^{\prime },q^{\prime }}\delta_{s,q}.
\end{eqnarray}
Note that, the indices $s$ and $s^{\prime }$ are non-necessary equal so that each of the above operator products are responsible for the polariton annihilation in some mode $q$ and its creation in some other mode~$q'$ or in the same mode $q$ when $s'=s$. 

Combining all results we derive the matrix elements of the vibration Hamiltonian
\begin{widetext}
 \begin{multline}  \label{A171}
\bra{\bm\sigma',q',u'}\hat{H}_{vib}\ket{\bm\sigma,q,u}=-\braket{\bm\sigma'|%
\bm\sigma}\sum_{m,\mu}\frac{\hbar\Omega_\mu X_\mu}{N} ({\sigma_{m,\mu}^{%
\prime }}^*+\sigma_{m,\mu})e^{-\i(q^{\prime }-q) m}\big[\bm v_{q^{\prime },q}%
\big]_{u^{\prime },u} \\
+\braket{\bm\sigma'|\bm\sigma}\delta_{q^{\prime },q}\delta_{u^{\prime
},u}\hbar\sum_{m,\mu} \Omega_\mu{\sigma_{m,\mu}^{\prime }}^*\sigma_{m,\mu} +%
\braket{\bm\sigma'|\bm\sigma}\delta_{q^{\prime },q}\delta_{u^{\prime },u}
H_{Bath}({\bm\sigma^{\prime }}^*,\bm \sigma),
\end{multline}
where $\big[\bm v_{q^{\prime },q}\big]_{u^{\prime },u}$ denotes the $%
(u^{\prime },u)$ component of the matrix of the Hopfield coefficients $\bm %
v_{q^{\prime },q}$, 
\begin{equation} \label{vmatrix}
\bm v_{q^{\prime },q}=\left(%
\begin{array}{cc}
\big[\bm v_{q^{\prime },q}\big]_{1,1} & \big[\bm v_{q^{\prime },q}\big]_{1,0}
\\ 
\big[\bm v_{q^{\prime },q}\big]_{0,1} & \big[\bm v_{q^{\prime },q}\big]_{0,0}%
\end{array}%
\right)\equiv\left(%
\begin{array}{cc}
\sin\phi_{q^{\prime }}\sin\phi_q & -\sin\phi_{q^{\prime }}\cos\phi_q \\ 
-\cos\phi_{q^{\prime }}\sin\phi_q & \cos\phi_{q^{\prime }}\cos\phi_q%
\end{array}%
\right)
\end{equation}
Correspondingly we obtain 
\begin{multline}\label{A17}
\bra{\bm\sigma',q',u'}c_{m,\mu}\hat{H}_{vib}\ket{\bm\sigma,q,u}%
-\sigma_{m,\mu}\bra{\bm\sigma',q',u'}\hat{H}_{vib}\ket{\bm\sigma,q,u} \\
=-\braket{\bm\sigma'|\bm\sigma}\frac{\hbar\Omega_\mu X_\mu}{N}%
e^{-\i(q^{\prime }-q) m}\big[\bm v_{q^{\prime },q}\big]_{u^{\prime },u}+%
\braket{\bm\sigma'|\bm\sigma}\delta_{q^{\prime },q}\delta_{u^{\prime
},u}\hbar\Omega_\mu \sigma_{m,\mu}  
+\braket{\bm\sigma'|\bm\sigma}\delta_{q^{\prime },q}\delta_{u^{\prime },u} 
\frac{\partial}{\partial{\sigma_{m,\mu}^{\prime }}^*}H_{Bath}({\bm%
\sigma^{\prime }}^*,\bm \sigma).
\end{multline}
\end{widetext}

The eqs.~(\ref{eq1},~\ref{eq2}) after substitution of the results of eqs.~(%
\ref{A9},~\ref{A91}) and eqs.~(\ref{A17},~\ref{A171}) give rise to the
complete set of equations of motion for the polariton wavefunction. The
obtained equations, however, are overcomplicated and even their numeric
solution can be difficult. To proceed we appeal to the multiconfiguration
Hartree framework~\cite{MMC1992}. The multiconfiguration Hartree approach is the method restructuring the set of equations of motion
obtained in the time-dependent basis set. The rigorous approach to the
generic solution is discussed in~\cite{WG2019}. In our particular model
(eqs.~\ref{eq1},~\ref{eq2}) we end up with the equations similar to the
equations of the mean-field Hartree theory.

The eq.~(\ref{eq1}) contains integration over $\bm\sigma $ inside the window
cut in the narrow vicinity of~$\bm\sigma ^{\prime }$. Structurally the
equation can be presented as follows 
\begin{widetext}
\begin{equation}
\sum_{\bm\sigma }\braket{\bm\sigma'|\bm \sigma}\left[ \bar{S}(\bm\sigma
,q^{\prime },u^{\prime })+\sum_{m,\mu }\left( {\sigma _{m,\mu }^{\prime }}%
^{\ast }-\sigma _{m,\mu }^{\ast }\right) \Bar{\bar{S}}(m,\mu ;\bm\sigma
,q^{\prime },u^{\prime }).\right] =0,  \label{eq1stuc}
\end{equation}%
The explicit forms of the coefficients in eq.~(\ref{eq1stuc}) are 
{\small \begin{multline} \label{barS} 
\bar{S}(\bm\sigma ,q^{\prime },u^{\prime })\equiv -\dot{\mathcal{C}}(\bm%
\sigma ,q^{\prime },u^{\prime }|t)-\i \mathcal{C}(\bm\sigma ,q^{\prime
},u^{\prime }|t)\left( (1-u^{\prime })\Lambda _{-q}+u^{\prime }\Lambda
_{+q}\right) 
+\i \sum_{q,u}\sum_{m,\mu }\frac{\Omega _{\mu }X_{\mu }}{N}(\sigma
_{m,\mu }^{\ast }+\sigma _{m,\mu })e^{-\i (q^{\prime }-q)m}\big[\bm %
v_{q^{\prime },q}\big]_{u^{\prime },u}\mathcal{C}(\bm\sigma ,q,u|t) \\
-\frac{\i }{\hbar }\mathcal{C}(\bm\sigma ,q^{\prime },u^{\prime
}|t)H_{Bath}(\bm\sigma ^{\ast },\bm\sigma )-\i \mathcal{C}(\bm\sigma
,q^{\prime },u^{\prime }|t)\sum_{m,\mu }\Omega _{\mu }\sigma _{m,\mu }^{\ast
}\sigma _{m,\mu } 
+\frac{1}{2}\mathcal{C}(\bm\sigma ,q^{\prime },u^{\prime }|t)\sum_{m,\mu }%
\left[ \sigma _{m,\mu }\dot{\sigma}_{m,\mu }^{\ast }-\sigma _{m,\mu }^{\ast }%
\dot{\sigma}_{m,\mu }\right] ;
\end{multline}}
\begin{multline}
\Bar{\bar{S}}(m,\mu ;\bm\sigma ,q^{\prime },u^{\prime })\equiv \i
\sum_{q,u}\frac{\Omega _{\mu }X_{\mu }}{N}e^{-\i (q^{\prime }-q)m}\big[%
\bm v_{q^{\prime },q}\big]_{u^{\prime },u}\mathcal{C}(\bm\sigma ,q,u|t)-%
\mathcal{C}(\bm\sigma ,q^{\prime },u^{\prime }|t)\dot{\sigma}_{m,\mu }
\label{barbarS} \\
-\i \mathcal{C}(\bm\sigma ,q^{\prime },u^{\prime }|t)\Omega _{\mu
}\sigma _{m,\mu }-\frac{\i }{\hbar }\mathcal{C}(\bm\sigma ,q^{\prime
},u^{\prime }|t)\frac{\partial }{\partial {\sigma ^{\prime }}_{m,\mu }^{\ast
}}H_{Bath}({\bm\sigma ^{\prime }}^{\ast },\bm\sigma ).
\end{multline}%
The structure of the second equation eq.~(\ref{eq2}) is similar to that of
eq.~(\ref{eq1stuc}) up to the integration measure. The equation in terms of
the above introduced functions (eq.~\ref{barS},~\ref{barbarS}) reads 
\begin{multline}
\sum_{\bm\sigma }\braket{\bm\sigma'|\bm \sigma}\mathcal{C}^{\ast }(\bm\sigma',q',u'|t)  \label{eq2stuc}  \left[ \Bar{\bar{S}}(m,\mu ;\bm\sigma ,q',u')+\sigma _{m,\mu }\bar{S}(\bm\sigma ,q',u')+\sigma
_{m,\mu }\sum_{n,\nu }\left( {\sigma _{n,\nu }'}^{\ast }-\sigma
_{n,\nu }^{\ast }\right) \Bar{\bar{S}}(n,\nu ;\bm\sigma ,q',u')\right] =0.
\end{multline}
\end{widetext}

The integration domain in eqs.~(\ref{eq1stuc}),~(\ref{eq2stuc}) is defined bythe scalar product of the coherent states, $\braket{\bm\sigma'|\bm \sigma}=\exp\left(-\sum_{m,\mu}\left\vert \sigma_{m,\mu}'-\sigma_{m,\mu}\right\vert^2/2\right)$. Due to the large total number of
the oscillators this product cuts a very narrow region in the whole configuration space. In the semiclassical approximation we set $\braket{\bm\sigma'|\bm \sigma}=\delta(\bm\sigma'-\bm\sigma)$, so that the resulting set of equations read $\bar{S}(\bm \sigma,q',u')=0$ and $\Bar{\bar{S}}(m,\mu;\bm \sigma, q',u')=0$. The same set of equations can be obtained from the assumption that the terms $\bar{S}$ and $\Bar{\bar{S}}$ nullify independently. Remarkably, this
solution exhausts all possible solutions. Here is the sketch of the proof. First, we formally convert the integral eqs.~(\ref{eq1stuc},~\ref{eq2stuc}) into a matrix form. For that we are indexing by integer numbers the nonintersecting domains of $\bm\sigma$ and replace the integration by summation. Also we enumerate all possible couples of the vibration indexes $(m,\mu)$ and of the polariton states $(q',u')$. The matrix of the coefficients of the obtained system of linear equations (the variables are $\bar{S}$ and $\Bar{\bar{S}}$) is quadratic and, generically, its determinant is non-zero. This immediately means that only trivial solution satisfies the equations. The determinant can become zero if one or several coefficients $\mathcal{C}(\bm \sigma,q',u'|t)$ are equal zero. 
Obviously, when all coefficients $\mathcal{C}(\bm \sigma,q',u'|t)$ are equal to zero there is no polariton excited and the
vibration degrees of freedom behave independently, which problem is out of
our attention. Now we show that the solution with the non-zero $\mathcal{C}(\bm \sigma,q',u'|t)$ is self-consistent solution of the
system. By equating $\bar{S}(\bm \sigma,q',u')=0$ we
obtain the first equations of motion,
\begin{widetext}
\begin{equation} \label{Csigmadot}
\mathcal{C} (\bm \sigma,q',u'|t) \left[ \dot{\sigma}%
_{m,\mu}+\i\Omega_\mu \sigma_{m,\mu}+\frac{\i}{\hbar} \frac{\partial}{%
\partial {\sigma^{\prime }}_{m,\mu}^*} H_{Bath}({\bm\sigma^{\prime }}^*,\bm %
\sigma)\right] 
= \i \frac{\Omega_\mu X_\mu}{N}\sum_{q,u} e^{-\i(q^{\prime }-q) m}\big[\bm %
v_{q^{\prime },q}\big]_{u^{\prime },u}\mathcal{C} (\bm \sigma,q,u|t).
\end{equation} 
The other condition $\Bar{\bar{S}}(m,\mu;\bm \sigma, q',u')=0$ yields the second equation of motion, 
\begin{multline}  \label{Cdot}
\dot{\mathcal{C}}(\bm \sigma,q',u'|t)=- \i \mathcal{C} (
\bm \sigma,q',u'|t)\left((1-u')\Lambda_{-q}+u'\Lambda_{+q}\right)  
+\i\sum_{q,u}\sum_{m,\mu}\frac{\Omega_\mu X_\mu}{N} (\sigma_{m,\mu}^*+%
\sigma_{m,\mu})e^{-\i(q^{\prime }-q) m}\big[\bm v_{q^{\prime },q}\big]%
_{u^{\prime },u}\mathcal{C} (\bm \sigma,q,u|t) \\
-\frac{\i}{\hbar}\mathcal{C} (\bm \sigma,q^{\prime },u^{\prime }|t) H_{Bath}(%
\bm\sigma^*,\bm \sigma)-\i\mathcal{C} (\bm \sigma,q^{\prime },u^{\prime
}|t)\sum_{m,\mu} \Omega_\mu \sigma_{m,\mu}^*\sigma_{m,\mu}  
+\frac{1}{2} \mathcal{C} (\bm \sigma,q^{\prime },u^{\prime }|t) \sum_{m,\mu}
[\sigma_{m,\mu} \dot{\sigma}_{m,\mu}^*- \sigma_{m,\mu}^*\dot{\sigma}%
_{m,\mu}].
\end{multline}

Consider the first consequence of the formulated equations (eqs.~\ref{Csigmadot},~\ref{Cdot}): a special combination of the equations, which generates the equation for the time-evolution of the squared amplitude $\abs{
\mathcal{C} (\bm \sigma,q^{\prime },u^{\prime }|t)}^2$, yields 
\begin{multline}  \label{ballanceC}
\frac{\partial}{\partial t}\abs{\mathcal{C} (\bm \sigma,q',u'|t)}^2 = \i \sum_{q,u}\big[\bm v_{q',q}\big]_{u',u}  \Big[ \mathcal{C}(\bm \sigma,q',u'|t)\alpha^*(q'-q )
\mathcal{C}^* (\bm \sigma,q,u|t)- \mathcal{C}^* (\bm \sigma,q',u'|t) \alpha(q'-q ) \mathcal{C} (\bm \sigma,q,u|t)\Big]
\end{multline}\end{widetext}
with the electron-vibration coupling $\alpha(q'-q)$ given by the
expression 
\begin{eqnarray}\label{transmissionprob}
\alpha(q^{\prime }-q )&\equiv& \frac{1}{N}\sum_{m} \alpha_m e^{-\i(q^{\prime
}-q) m},\\ \alpha_m &\equiv& \sum_{ \mu} \Omega_\mu X_\mu
(\sigma_{m,\mu}^*+\sigma_{m,\mu}).
\end{eqnarray}  
At the derivation of eq.~(\ref{ballanceC}) it has been assumed that the bath Hamiltonian is linear in $%
\sigma_{m,\mu}$, which is the standard form of the phonon interaction
Hamiltonian. The obtained equation (eq.~\ref{ballanceC}), in particular,
shows that the amplitudes $\left\vert \mathcal{C} (\bm \sigma,q',u'|t)\right\vert$ for all allowed values of $q$ get non-zero
values as soon as any of $\alpha_m$ has a non-zero value. This finishes the
proof that the non-zero coefficients $\mathcal{C}(\bm \sigma,q',u'|t)$ generate a self-consistent solution of the system eqs.~(\ref{eq1stuc}), and~(\ref{eq2stuc}).

One more critical condition follows from eq.~(\ref{ballanceC}), it is conservation of the total probability of finding polariton in any of the allowed states. Indeed, since $\alpha^*(q'-q)=\alpha(q-q')$ and due to the symmetry of the matrix $v_{q,q'}$ ($[v_{q,q'}]_{u,u'}=[v_{q',q}]_{u',u}$) after summation over $q'$ and $u'$ we immediately derive 
\begin{equation}
\dot{\mathcal{N}}(t|\bm\sigma)=0,\quad \mathcal{N}(t|\bm\sigma)=\sum_{q,u}
\left\vert \mathcal{C} (\bm \sigma,q,u|t)\right\vert^2.
\end{equation}

The total probability conservation allows us to sum up the equations~(\ref{Csigmadot}) over $q'$ and $u'$ after multiplication of each of them by the corresponding coefficient $\mathcal{C}^* (\bm \sigma,q',u'|t)$. In such a way we derive the equation of motion for the $(m,\mu)$ oscillator 
\begin{multline}\label{sigmadot}
\dot{\sigma}_{m,\mu} =-\i\Omega_\mu \sigma_{m,\mu}+\i\frac{\Omega_\mu X_\mu}{N}\chi_m(t|\bm\sigma) \\
-\frac{\i}{\hbar}\frac{\partial}{\partial{
\sigma_{m,\mu}'}^*}H_{Bath}({\bm\sigma'}^*,\bm \sigma),
\end{multline}  
This equation contains the mean-field Hartree term $\i\frac{\Omega_\mu X_\mu}{N}\chi_m(t|\bm\sigma)$, which describes influence of the polariton field on the quantum oscillators. The Hartree term amplitude is 
\begin{equation}  \label{chi0}
\chi_m(t|\bm\sigma)= \sum_{q',q,u',u}e^{-\i(q'-q)
m}\big[\bm v_{q',q}\big]_{u',u}\rho\left(t\left |\atop{q,u}{q',u'}\right|\bm \sigma\right)
\end{equation}
where $\rho$ is the density matrix, 
\begin{equation}  \label{rho}
\rho\left(t\left |\atop{q,u}{q',u'}\right|\bm \sigma\right)=\frac{\mathcal{C}^*(\bm \sigma,q',u'|t) 
\mathcal{C}(\bm \sigma,q,u|t)}{\mathcal{N}(t|\bm\sigma)}.
\end{equation}
The symmetry $[v_{q,q'}]_{u,u^{\prime }}=[v_{q',q}]_{u',u}$ guaranties that $\chi_m$ is a real-valued function,
i.e. $\chi_m^*(t|\bm\sigma)=\chi_m(t|\bm\sigma)$.

At the final step we substitute $\dot{\sigma}_{m,\mu}$ from eq.~(\ref{sigmadot}) into eq.~(\ref{Cdot}) to obtain the second equation of motion 
\begin{multline}  \label{Cdot1}
\dot{\mathcal{C}}(\bm \sigma,q',u'|t)=- \i
\left((1-u')\Lambda_{-q'}+u'\Lambda_{+q'}\right)\mathcal{C} (\bm \sigma,q^{\prime },u'|t) \\
-\frac{\i}{2N} \mathcal{C} (\bm \sigma,q',u'|t)
\sum_{m}\alpha_m \chi_m(t|\bm\sigma)\\ + \i \sum_{q,u} \alpha(q'-q )
\big[\bm v_{q',q}\big]_{u',u}\mathcal{C} (\bm \sigma,q,u|t).
\end{multline}

Eq.~(\ref{Cdot1}) together with eq.~(\ref{sigmadot}) form a system of equations of vibration assisted polariton motion. Deriving the equations we did not make any approximations.

To investigate influence of the Hartree term in the next section we consider two problems: we calculate the polariton luminescence using the diagonal approximation in the equations of motion, sec.~\ref{Sec32}; and solve the equations of motion in the vicinity of AP, sec.~\ref{sec21}.

\section{Solution of polariton equations in various regimes.}
\label{Sec2}
It is instructive to investigate the influence of the nonlinear Hartree term in a somewhat simplified setup. To this end, we consider two types of problems, namely we calculate the fluorescence spectra in the regime of large Rabi splitting when some sort of linearisation of the equations of motion is possible (section~\ref{Sec32}). Second, in the section~\ref{sec21} we consider the evolution of the polariton wavefunction in the AP vicinity, when nonlinearity plays a crucial role.

\subsection{Fluorescent spectrum in quasi-diagonal approximation}

\label{Sec32}

\subsubsection{Equations of motion at large Rabi splitting in the quasi-diagonal approximation}

\label{Sec321} In the quasi-diagonal approximation one  keeps only the diagonal terms, $q'=q$, corresponding to the largest value of the vibational perturbation~\cite{Fainberg22JPCA} in the second equation of motion (eq.~\ref{Cdot1}). The processes with $q\ne q'$ describe the polariton relaxation along the dispersion curve. In our consideration we neglect such processes assuming that they are much slower than any other process under consideration, so we make use the replacement $\alpha (q'-q)\rightarrow \alpha (0)$ in eq.~(\ref{Cdot1}). We return back to the discussion of this issue at the end of the section~\ref{Sec312}. In addition we consider the regime of the large Rabi splitting, i.e. when the difference  $\Lambda_{-q}-\Lambda_{+q}$ is larger of all relevant charateristic frequencies of the problem. In this case the terms describing the exchange between the polariton branches can be neglected. Using explicit form of the diagonal entries of the matrix $\bm v$ (eq.~\ref{vmatrix}), which are the electronic Hopfield coefficients $\sin^2\phi_q$ and $\cos^2\phi_q$ corresponding to the upper and lower polariton branches, respectively, we obtain 
\begin{widetext}
\begin{eqnarray}
  \dot{\mathcal{C}}(\bm\sigma ,q,0|t) &=&-\i\left(\Lambda _{-q} 
-2\alpha (0)\Bigg[\cos^2\phi_q  
-\frac{1}{2}\sum_{q}\left( \sin^2\phi_q \abs{\mathcal{C}(\bm 
\sigma ,q,1|t)}^2 + \cos^2 \phi_q \abs{ \mathcal{C}(\bm \sigma ,q,0|t)}^2 \right) \Bigg]\right)\mathcal{C}(\bm\sigma ,q,0|t); 
 \label{diagC0}\\
 \label{diagC1} 
 \dot{\mathcal{C}}(\bm\sigma ,q,1|t) &=&-\i \left(\Lambda _{+q} 
-2\alpha (0)\Bigg[\sin ^{2}\phi _{q} -\frac{1}{2}\sum_{q}\left( \sin ^{2}\phi _{q}\left\vert \mathcal{C}(\bm%
\sigma ,q,1|t)\right\vert ^{2}+\cos ^{2}\phi _{q}\abs{ \mathcal{C}(\bm \sigma ,q,0|t)}^2 \right) \Bigg]\right)\mathcal{C}(\bm\sigma ,q,1|t). 
\end{eqnarray}
\end{widetext}
The factor 2 at $\alpha (0)$ appears due to the symmetry $q\to -q$. The factors with the opposite momenta contribute identically. 

To estimate the time dependence of $\alpha (0)$ we return back to eq.~(\ref{sigmadot}). Instead of solving all equations for $\sigma _{m,\mu }$ independently we replace the Hartree $m$-dependent term, $\chi_{m}$ (eqs.~\ref{chi0},~\ref{rho}), by its average
value, an $m$-independent approximate polariton field acting on each molecule identically. For that we use the diagonal approximation and omit the highly oscillating terms in the same way as we did in the solutions eqs.~(\ref{diagC0}), and ~(\ref{diagC1}). These approximations are identical to the ``maximal action'' approximation, when one replace $\chi _{m}$  by the exact upper border of its estimator, see Appendix~\ref{AppD}. Therefore the approximate equations for the function $\sigma_{m,\mu}$ become 
{\small \begin{multline}\label{xmmu}
\dot{\sigma}_{m,\mu }\approx -\i \Omega _{\mu }\sigma _{m,\mu }-2\i \gamma \mathrm{Im}\sigma_{m,\mu }-\i \xi _{m}(t)+\i \Omega _{\mu }X_{\mu }\\ \times\sum_{q}\left( \sin ^{2}\phi _{q}\left\vert \mathcal{C}(\bm\sigma ,q,1|t)\right\vert ^{2}+\cos ^{2}\phi _{q}\left\vert \mathcal{C}(\bm\sigma ,q,0|t)\right\vert ^{2}\right).  
\end{multline}}
Here we modelled the bath degrees of freedom by some damping with the
rate $\gamma $ and by a stochastic force $\xi _{m}(t)$. 
Eq.~(\ref{xmmu}) is the Langevin equation with an external force. Following the standard procedure we assume that the noise $\xi _{m}(t)$ is Gaussian with zero mean value. We imply that the noise is $\delta $-correlated in time, i.e. $<\xi _{m}(t)\xi _{n}(\tau )>=\gamma \Omega _{\mu }k_{B}T (\delta_{m,n}+R_{m,n})\delta (t-\tau )$ ($k_{B}T$ is the bath temperature expressed in the energy units) and $R_{m,n}$ generates some (small) correlations between the molecules. We make the stochastic averaging in the regime when the equilibration of the vibrations happens very fast after the optical excitation. For the case of overdamped oscillator the averaged over the noise function $\sigma _{m,\mu }$ is easy to calculate, it is 
{\small \begin{multline}\label{xmmu1}
<\sum_{m}(\sigma _{m,\mu }^* +\sigma _{m,\mu })>=2NX_{\mu
}\\\times \sum_q \left( \sin^2\phi_q\abs{\mathcal{C}(\bm\sigma
,q,1|t)}^2+\cos^2\phi_q\abs{\mathcal{C}(\bm\sigma ,q,0|t)}^2\right),
\end{multline}}
which means that the averaged $<\alpha (0)>$ is proportional to the Hartree term, 
\begin{multline}\label{alphamean}
<\alpha (0)>\equiv <\frac{1}{N}\sum_{m}\alpha_m>=\omega _{st}
\sum_{q}\bigg(
\sin ^{2}\phi _{q}\\\times \abs{\mathcal{C}(\bm\sigma ,q,1|t)}^2 +\cos^2\phi_q\abs{ \mathcal{C}(\bm\sigma ,q,0|t)}^2\bigg).  
\end{multline}
Reformulation of the stochastic equation (eq.~\ref{xmmu}) in terms of the
Fokker-Planck equation and consequent application of the theorem for a sums
of weighted normally distributed random variables allows us to write down
the distribution for $\alpha (0)$ as \cite{Fainberg22JPCA} 
\begin{equation}\label{Probmeasure}
P\big(\alpha (0)\big)=\sqrt{\frac{1}{2\pi k_{N}\omega _{st}k_{B}T}}\; e^{ 
 -\frac{(\alpha (0)-<\alpha (0)>)^{2}}{2k_{N}\omega _{st}k_{B}T}},
\end{equation}
where
\begin{equation}
k_{N}=(N\mathcal{+}2\sum_{i<j}^{N}r_{ij})/N^{2},  \label{eq:k_N}
\end{equation}%
$0\leqslant r_{ij}\leqslant 1$ are the correlation coefficients that are
different from zero when the vibrations include both the intra- and the
intermolecular ones, see the definition of the stochastic force $\xi_m(t)$ in the paragraph between eqs.~(\ref{xmmu}), and~(\ref{xmmu1}). The assumption of the intramolecular nature of the optically active vibrations means that in this case, they are statistically independent, so that the coefficient $k_{N}$ equals $1/N$. In the other extreme case when the optically active vibrations are intermolecular ones, the correlation coefficients $r_{ij}=1$, and $k_{N}=1$.

\subsubsection{Luminescence spectrum calculation}\label{Sec312}
In an experiment, the polariton system is irradiated by the pumping light and emits light which carries information about the polariton states. The pronounced advantage of the polariton devices is the one-to-one correspondence between the polariton states and the emitted photons. Since the emitted photon is a part of the polariton particle it preserves the polariton energy and the in-plane wave vector. A fixed polariton decay rate makes it possible studying of the dynamics of these composite particles.

\begin{figure}
\begin{tabular}{lc}
a)&\includegraphics[scale=0.3]{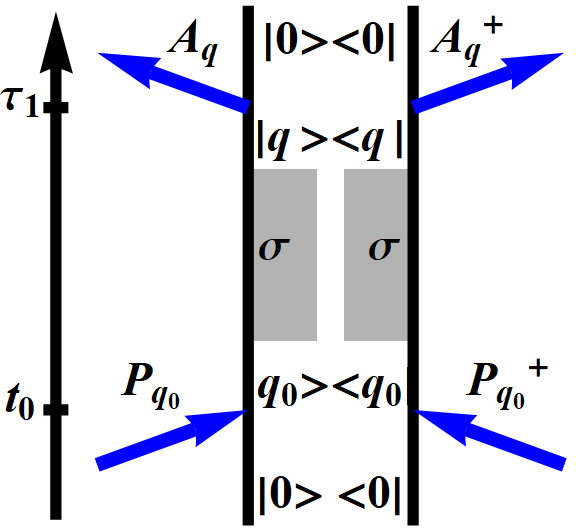}\vspace{5pt}\\
b)&\includegraphics[scale=0.37]{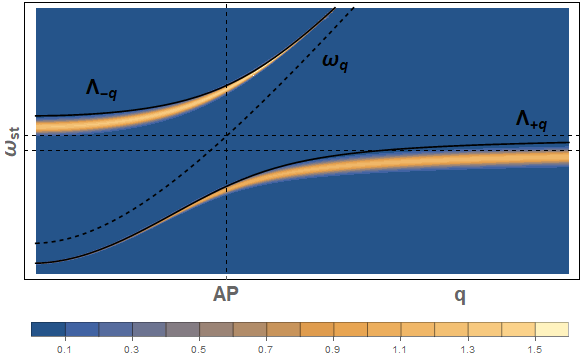}
\end{tabular}
\caption{ \small a) The ladder-like diagram depicting the contribution to the fluorescent signal (eq.~\protect\ref{factorization}): two polaritons (vertical lines), initially (at time $t_0$) created in the conjugated states with the momentum $q_0$ (operators $P_{q_0}$ and $P_{q_0}^\dag$), evolve due to interaction with the vibrational bath (the region denoted by $\bm\protect \sigma$) to the states with the momentum $q$ and decay into intracavity photons (operators $A_q$ and $A_q^\dag$). The system's initial and final states are the vacuum states $\ket{0}$. b) Pictorial representation of the cavity polariton fluorescence intensity (in arbitrary units) plotted v.s. the luminescence frequency $\omega$ and the wavevector $q$  in the assumption that the polariton is excited at a single $q$, i.e. $\mathcal{F}_0(q')=\frac{1}{\sqrt{2}}\left( \delta _{q,q'}+\delta_{q,-q'}\right) $ and $\mathcal{F}_1(q')=0$ in eq.~(\ref{signal}) for the lower polariton branch when the radiation is measured at the same $q$. The upper polariton branch is plotted by analogy. The dispersion curves of the cavity photons $\omega_q$ and the polariton dispersion curves, $\Lambda_{\pm q}$, are shown for comparison. In the numerics we used the parameters $\omega_0=3.1$eV, $\omega_{st}=0.15$eV, $g=0.5$eV, $k_N k_BT=0.03$eV, the dispersions $\Lambda_{\pm q}$ (eq.~\ref{Lambd}) are calculated with the substitution $\omega_{ex}=\omega_0+\omega_{st}/2$.}\label{FigDiagr}
\end{figure}

Formally, to relate the outer field with the intracavity one, we appeal to the quasimode approximation~\cite{SPQST1999}, when the in-out coupling conserves the in-plane components $q_{||}$ of the intracavity wavevector $q=(q_{||},q_z)$ (see fig.~\ref{fig2a}). The $z$-components of the wavevector in free space are determined by the emitted photon energy. Thus to denote the external electromagnetic field components (emitted photons) one can use solely the wavevector component $q_{||}$~\cite{Zoubi_Rocca05,Lidzey08,Rocca09}. The luminescence signal amplitude $S_{q_{||}}(\omega )$ detected out of the sample and coming from the direction marked by the wavevector $q_{||}$, which frequency $\omega _{\tilde{q}}=\frac{c}{n_{0}}\abs{\tilde{q}}$  is generally calculated from the two-time correlation function of the quantized electric field generated by the leakage of photons (with the characteristic rate $\kappa $) through the mirrors of the microcavity~\cite{EW1981}. The details of the calculation are given in appendix~\ref{AppB}. The signal amplitude $S_{q_{||}}(\omega )$ is expressed in terms of the Fourier transform of the quantum correlation functions $G_{1}(\omega ,\omega _{3})$ and $G_{2}(\omega ,\omega _{3})$ (eqs.~\ref{G1AppB},~\ref{G2AppB}), 
\begin{multline}
S_{q_{||}}(\omega )\propto \hbar \omega \frac{\bar{r}_{ex}^2\kappa^2}{\pi }\mathrm{Re}\int_{-\infty }^{+\infty }d\omega_3 \\ 
<\Big(G_{1}(-\omega
,\omega _{3})G_{2}(\omega ,-\omega _{3})\\+G_{1}^{\ast }(\omega ,\omega
_{3})G_{2}^{\ast }(-\omega ,-\omega _{3})\Big)>,
\end{multline}
where the frequency $\omega$ coincides with the frequency of the wave
freely propagating in the outer space. 
The coefficients $\kappa$ and $\bar{r}_{ex}$ are the effective rates of the polariton decay and creation, respectively. In addition to the quantum average of the correlation functions, we also perform the thermodynamic average, which is denoted by $<\dots >$.
The quantum correlation function $\bra{0}P_{q_{0}}Te^{\i
\int_{t_{0}}^{\tau _{1}}\hat{\mathcal{H}}(\tau )d\tau /\hbar }A_{q}^{\dag }\ket{0}$, entering the functions $G_{1}(\omega ,\omega _{3})$ and $%
G_{2}(\omega ,\omega _{3})$ (eqs.~\ref{G1AppB},~\ref{G2AppB}), describes
evolution of the wave-function of a polariton created with the wavevector $q_0$ at some instant of time $t_{0}$. The polariton evolves up to the time $\tau _{1}$ when it is annihilated at the state with the wavevector $q$. Schematically the process is depicted by the diagram shown in fig.~\ref{FigDiagr}. The equations connecting the material and the polariton
operators (eqs.~\ref{Pop},~\ref{Qop}) allows us to express the field
operator $A_{q}$ in terms of the operators $P_{q}$ and $Q_{q}$. Assuming, at the moment, that only the lower branch is excited we use the replacement $A_{q}\rightarrow \sin \phi _{q}P_{q}$. Calculation of the spectrum essentially depends on the particularities of the polariton creation process. To specify it, we focus on the process when the polaritons is created by a short light pulse at the instant of time $t_{0}$ with the initial distribution $\mathcal{F}_{u}(q)$ (the subscript $u$ stands for the lower, $u=0$ and the upper $u=1$ branch). Therefore, after simplification the expression for the signal reduces to 
\begin{multline}\label{Sqpar}
S_{q_{||}}(\omega )\propto \hbar \omega \bar{r}_{ex}^2\kappa^2 <G(\omega
)G^*(\omega )>,\\ G(\omega )=\int_{t_0}^{+\infty }d\tau
_1 e^{\i \omega \tau_1}\mathcal{C}(\bm\sigma ,q,0|\tau _1),
\end{multline}
where $\mathcal{C}(\bm\sigma ,q,0|\tau _{1})$ is the solution of the
polariton equations of motion with the initial condition $\mathcal{C}(\bm 
\sigma ,q,u|t_{0})=\mathcal{F}_{u}(q)$. The factor $\sin^2\phi_q$ in eq.~(\ref{Sqpar}) was included into the parameters $\bar{r}_{ex}^2\kappa^2$.

In the diagonal approximation the evolution of the polariton expansion coefficient is defined by eqs.~(\ref{diagC0}), and~(\ref{diagC1}). From these equations, it follows immediately that the amplitude of the wavefunction is conserved, while the time dependence is contained solely in the phase factor. Thus, the formal solution is given by 
\begin{equation}
\mathcal{C}(\bm\sigma ,q,0|t)=\mathcal{F}_{0}(q)e^{ -\i \Lambda
_{-q}(t-t_{0})+\i \left\vert u_{q}\right\vert ^{2}\int_{t_{0}}^{t}\alpha
(0|\tau )d\tau } 
\end{equation}%
where for convenience we introduced the notation for the effective lower branch Hopfield coefficient 
\begin{equation}
\abs{ u_q}^2\equiv 2\left[ \cos^2\phi _{q}-\frac{1}{2}
\sum_{q'}\cos^2\phi _{q'} \mathcal{F}
_0^2(q') \right]   \label{uqmean}
\end{equation}
Therefore in the regime of thermal equilibrium of the vibration subsystem the function $G(\omega )$ can be immediately calculated to give 
\begin{equation}\label{GreenF}
G(\omega )=\frac{\i e^{\i \omega t_{0}}\mathcal{F}_{0}(q)}{\omega
-\Lambda _{-q}+\abs{u_q}^2\alpha (0)+\i \bar{\gamma}},  
\end{equation}
where the small parameter $\bar{\gamma}$ is introduced for regularisation of the integrals. After averaging with the probability measure eq.~(\ref{Probmeasure}) we eventually derive 
\begin{multline}\label{signal}
S_{q_{||}}(\omega )\propto \sqrt{\frac{\pi }{2k_{N}\omega _{st}k_{B}T}}\frac{\hbar \omega \bar{r}_{ex}^{2}\kappa ^{2}\mathcal{F}_{0}^{2}(q)}{\abs{u_q}^2}\\\times\exp \left[ -\frac{\left( \omega -\Lambda
_{-q}+<\alpha (0)>\abs{u_q}^2\right)^2}{2k_{N}\abs{u_q}^4\omega_{st}k_B T}\right],
\end{multline}

In the final expression for the fluorescent signal $S_{q_{||}}(\omega )$ (eq.~\ref{signal}) shows that the frequency is distributed around the polariton frequency $\Lambda _{-q}$, which is shifted by the factor $-<\alpha (0)>\abs{u_q}^2$. In the quasi-stationary regime (i.e. when the polariton relaxation process is much slower than any other proses, see the first paragraph of the section~\ref{Sec321}) we can approximate the polariton time-dependent wavefunction coefficients in eq.~(\ref{alphamean}) by their initial values, so using the normalizing condition for $\mathcal{F}_u(q)$ we write
\begin{multline}\label{alpha0averaged}
<\alpha (0)>=\omega _{st}\\-\omega _{st}\sum_{q}\left( \cos ^{2}\phi
_{q}\left\vert \mathcal{F}_{1}(q)\right\vert ^{2}+\sin ^{2}\phi
_{q}\left\vert \mathcal{F}_{0}(q)\right\vert ^{2}\right).
\end{multline}
Therefore both the frequency shift and the distribution width depend on the initial population of the mode $q$,  $\mathcal{F}_{0}(q)$. Eq.~(\ref{signal}) structurally reproduces the result eq.~(72) in~\cite{Fainberg22JPCA}, where the factor $\omega _{st}\left\vert \bar{u}_{s0}(\bm q)\right\vert ^{2}$ has to be replaced now by the product $<\alpha (0)>\abs{u_{q}}^2$. Therefore, our rigorous approach shows that the spectral shift is produced by the Hartree term. Notice also that the luminescence spectrum (eq.~\ref{signal}) is narrowing for the intramolecular nature of the low frequency optically active vibrations ($k_{N}=1/N$) as the number of molecules $N$ increases. The narrowing of the polariton luminescence spectrum by increasing the number of molecules was predicted in Ref.~\cite{Fainberg22JPCA} and resembles the exchange (motional) narrowing in the absorption of molecular aggregates \cite{Knapp84}. The difference lies in the nature of the interaction responsible for the exchange effects~\cite{Fainberg22JPCA}. 

In fig.~\ref{FigDiagr} b we plotted the numeric estimation of the polariton fluorescence for all values of $q$ according to the obtained formulae (eqs.~\ref{uqmean},~\ref{signal},~\ref{alpha0averaged}). We also extended our numerics for the upper polariton branch. It is assumed that the fluorescence is measured at the same wavevector $q$ where the polariton was excited. The two limiting photonic and excitonic regimes are smoothly connected in the intermediate region.

For completeness, we investigate the obtained solution in two limiting cases. The simplest for the analysis case is the one which describes the polaritons created with a small momentum $q$ and the initial amplitude $\mathcal{F}_{0}(q')=\frac{1}{\sqrt{2}}\left( \delta _{q,q'}+\delta_{q,-q'}\right)$. In this case the Hopfield coefficient $\cos^2\phi_q$ is very small. Thus the combination of factors generating the frequency shift, $-<\alpha (0)>\abs{u_q}^2\approx -\omega _{st}\cos^4\phi_q$ (eqs.~\ref{alphamean},~\ref{uqmean}), is essentially suppressed by the fourth order of $\cos \phi_q$ and can be neglected. In the limit when $\cos^2\phi _q\rightarrow 0$ the spectrum formula can be further reduced by using the $\delta $-function Gaussian representation, $\delta (x)=\lim_{\beta \rightarrow 0}\frac{1}{\beta \sqrt{\pi }}e^{-\left( \frac{x}{\beta }\right)^2}$. Since in the case under consideration $\Lambda _{-q}\sim \omega _{q}$ (see fig.~\ref{fig2a} b), the system radiates at the photon frequency, i.e. $S_{q_{||}}(\omega )\propto \hbar \omega \delta (\omega -\omega_q)$, see the upper polariton branch at large $q$ and the lower branch for small $q$ on the fluorescence plot in fig.~\ref{FigDiagr} b. 

When the polariton is created in the state with large $q$, the lower plariton branch Hopfield coefficient $\cos ^{2}\phi _{q}$ becomes very close to unity, so that the second term in eq.~(\ref{alpha0averaged}) can be neglected. When the polariton field is initially fully concentrated at the wavevector $q$ so that $\abs{u_q}^2\approx 1$. The expansion of $\Lambda _{-q}$ over small $\omega _{ex}/\omega _{q}$ and $g/\omega _{q}$ in the leading order gives $\Lambda _{-q}\approx \omega _{ex}$, so that the central luminescence frequency $\Omega $ coincides with the exciton radiation frequency $\Omega \approx \omega _{ex}-\omega _{st}=\omega_{0}+\omega _{st}/2-\omega _{st}=\omega _{0}-\omega _{st}/2$. The line width reaches the value $k_{N}\omega _{st}k_{B}T$, see the lower polariton branch at large $q$ and the upper branch for small $q$ on the fluorescence plot in fig.~\ref{FigDiagr} b. 

The diagonal approximation, which we use to obtain the spectrum (eq.~\ref{signal}), obviously, is not sufficient when the polariton characteristic decay time is large in comparison with the effective inverse rate of transitions between the states with various $q$. During this time the polariton wavefunction amplitudes can spread over a large range of wavevectors. The time-dependence of the amplitudes $\abs{\mathcal{C}(\bm\sigma ,q,u|t)}^2$ and the change in the momentum distribution can influence the luminescence spectrum. To go beyond the diagonal approximation, one can additionally equip the problem (eqs.~\ref{diagC0},~\ref{diagC1}) by the system of balance equations, which are formulated for the density matrix $\rho(q,u|q',u';t)\equiv \mathcal{C}^{\ast }(\bm\sigma ,q',u'|t)\mathcal{C}(\bm\sigma ,q,u|t)$. Following the method proposed by Zwanzig~\cite{Z1964} von Neumann's equation can be resolved for the diagonal entries of the density matrix, $\rho (q,0|q,0;t)\equiv\left\vert \mathcal{C}(\bm\sigma ,q,0|t)\right\vert ^{2}$, (see details of derivation in the appendix~\ref{AppC}). The obtained balance equation (eqs.~\ref{balanceApp},~\ref{kernel}) describes the evolution of the polariton quantum amplitudes. Note that von Neumann's equation does not contain the Hartree term in any explicit form. The estimation of the transition rate in the leading order yields  $\sqrt{k_N\omega_{st}k_BT/N}$ (see appendix~\ref{AppC}). The small factor $\sqrt{k_N/N}$, suppressing the polariton relaxation along the dispersion curve, gives us the supporting argument in favour of our quasi-diagonal approximation used for calculation of the fluorescent spectrum (eq.~\ref{signal}).

There are two more remarks. The obtained luminescence spectrum is defined as a thermal average of Green's function $G(\omega )$ (eq.~\ref{GreenF}). The pure polariton (no vibrations) Green's function pole is located at the polariton energy. The correction to its value ($\abs{ u_{q}}^2 \alpha (0)$) can be interpreted as the polariton self-energy. Indeed, its structure repeats the typical structure of the self-energy term, it is a product of the Hartree forces ($\chi _{m}(t|\bm\sigma )$) exciting the vibrations and those that enter the amplitude $\left\vert u_{q}\right\vert^{2}$ with the characteristic interaction energy $\omega _{st}$ (proportional to the oscillator strength). Also, we note that at the construction of our theory we made a voluntary decision, when inserted the $\omega _{st}/2$ directly into the definition of the polariton dispersion $\Lambda _{\pm q}$. This, however, can be done differently by inserting $\omega _{st}/2$ into the definition of $\alpha (q)$. This should not bring any difference when the problem is solved non-perturbatively, while the perturbative approach can be sensitive to this choice especially close to the AP.

To conclude this section we note that the effect of molecular Stokes shift on polariton spectra at a strong light-matter coupling was seen and discussed in a number of experimental works~\cite{LBVAS1999,TBKPN2017,HPTTGT2021}. It is worth noting that the theory developed in this paper and also in Ref.~\cite{Fainberg22JPCA} can serve as a basis for the heuristic model formulated in Ref.~\cite{TBKPN2017}. The latter model (see fig.~1 d in Ref.~\cite{TBKPN2017}) qualitatively explains the effect of the increase of the Stokes shift in the resonant cavities compared to the one measured for the same material (dye-doped films R6G:PMMA) deposited on glass. Indeed, according to eq.~(\ref{signal}) with the substitution eq.~(\ref{alpha0averaged}), the fluorescence signal maximum is found near the frequency $\Lambda _{-q}-\omega_{st}\abs{u_{q}}^2$, i.e. the polariton energy is corrected by the Stokes shift weighted with the excitonic contribution to the polariton. Moreover, our theory explains also the narrowing of the luminescence spectrum of R6G:PMMA film placed in the cavity with respect to the luminescence spectrum of the same film deposited on glass (see fig. 3 c in Ref.~\cite{TBKPN2017}). The second moment of the polariton luminescence spectral line equals to the second moment of the molecular luminescence ($\omega_{st}k_{B}T$) multiplied by the factor $k_{N}\abs{u_q}^4<1$ (eq.~\ref{signal}). This means that there are at least two sources of the spectral line narrowing observed in Ref.~\cite{TBKPN2017}: the motional narrowing (when$k_{N}<1$), and due to effective decrease of the excitonic component in the polariton accounted by the weight $\abs{u_q}^4<1$.

\subsection{Behaviour of the polariton wavefunction in the vicinity of AP}

\label{sec21}
In the previous section the solution of the equations of motion have been found in the regime where their linearisation is possible. In this section we consider evolution of the polariton wavefunction in the AP vicinity, when the nonlinear terms are essential. As a simplifying condition, we use the assumption that the polariton wavefunction is initially activated at a single momentum mode $q=q_0$. To formulate our equations for this case in a convenient way we introduce the following notations: $C_u\equiv \mathcal{C}(\bm \sigma,q_0,u|t)$, $s\equiv
\sin\phi_{q_0}$, $c\equiv \cos\phi_{q_0}$. We also redefine the time
variable as $t g\to t$, and set $\Lambda_\pm\equiv\Lambda_{\pm q_0}/g$. At
this choice the Rabi frequency becomes $(\Lambda_{+}-\Lambda_{-})=1/cs\ge2$. We also set $\mathcal{N}(t|\bm\sigma)=2\abs{C_1}^2+2\abs{C_2}^2=1$, then the Hartree factor simplifies to the $m$-independent function of time, $\chi_m(t|\bm \sigma)=2\abs{ sC_1-cC_0}^2$. The multiplier 2 arises from the identical contributions of the polariton modes with $+q_0$ and $-q_0$ due to the mirror symmetry of the dispersion curves. The equations for the components of $\bm \sigma$ no longer contain the $m$-dependence and we define the dimensionless frequency shift amplitude 
\begin{equation}\label{xoft}
x(t)=-\frac{1}{N g}\sum_{m}\alpha_m= -\frac{\alpha(0;t)}{g}.
\end{equation} 
In the case of $m$-independent vibrations the electron-vibration coupling $\alpha(q'-q)$ (eq.~\ref{transmissionprob}) is proportional to $N\delta_{q',q}$, so that there are no transitions between the states with various $q$, while the inter-branch exchange still takes place. In the above simplified notations, the equations of motion (eq.~\ref{Cdot1}) read 
\begin{eqnarray}  \label{C1singlemode}
\dot{C}_1&=&-\i \Lambda_{+} C_1+\i x(t) \abs{sC_1-cC_0}^2
C_1 \nonumber\\&&+2\i x(t) s \left( c C_0 -s C_1\right) \\
\dot{C}_0&=&-\i \Lambda_{-} C_0+\i x(t)\abs{sC_1-cC_0}^2 C_0\nonumber 
\\&& -2\i x(t) c \left(c C_0 - s C_1\right)  \label{C0singlemode}
\end{eqnarray}
It is convenient to use the equations written in a trigonometric form. For the following reparametrization of the polariton coefficients 
\begin{eqnarray}  \label{C1noiseless}
C_1(t)&=&\frac{1}{\sqrt{2}}e^{-\i\Lambda_+ t +\i
\Theta(t)+\i\eta(t)+\i\Phi(t) }\sin\Psi(t), \\
C_0(t)&=&\frac{1}{\sqrt{2}}e^{-\i\Lambda_- t +\i
\Theta(t)+\i\eta(t)-\i\Phi(t) }\cos\Psi(t).  \label{C0noiseless}
\end{eqnarray}
the set of equations for the coupled angles $\Psi$ and $\Phi$ reads 
\begin{eqnarray}
\dot{\Psi}(t)&=&2cs x(t)\sin\left(\frac{t}{cs}+\Phi(t)\right),
\label{dotPsi} \\
\dot{\Phi}(t)&=& (c^2-s^2)x(t)\nonumber\\&&+2cs x(t)\cot 2\Psi(t) \cos\left(\frac{t}{cs}%
+\Phi(t)\right).  \label{dotPhi}
\end{eqnarray}
The equations for the other two dependent angles are 
\begin{eqnarray}
\dot{\Theta}(t)&=& x(t)\big(2\abs{sC_1-cC_0}^2-1\big),
\label{dotTheta} \\
\dot{\eta}(t)&=&\frac{2 cs x(t)}{ \sin2\Psi(t)}\cos\left(\frac{t}{cs}
+\Phi(t)\right).  \label{doteta}
\end{eqnarray}

\begin{figure*} 
\begin{tabular}{cc}
\includegraphics[scale=0.38]{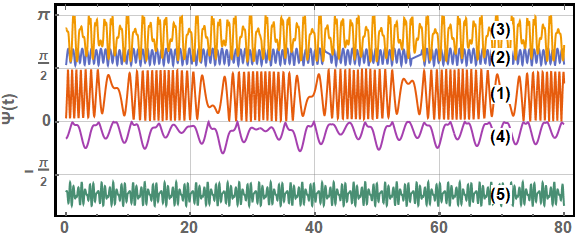} & %
\includegraphics[scale=0.38]{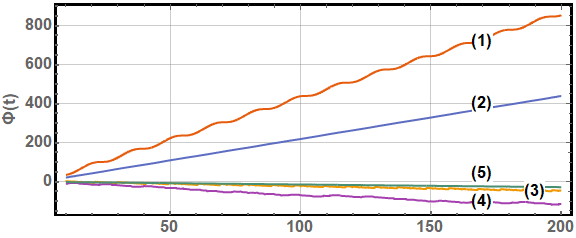} \\ 
\includegraphics[scale=0.385]{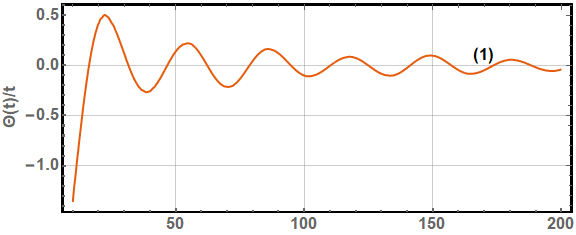} & %
\includegraphics[scale=0.38]{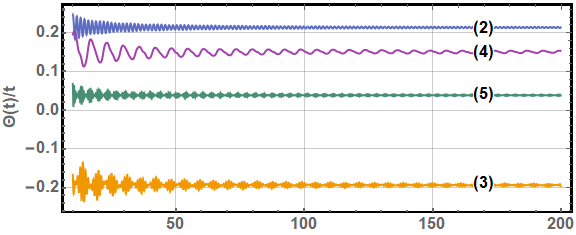} \\ 
\includegraphics[scale=0.375]{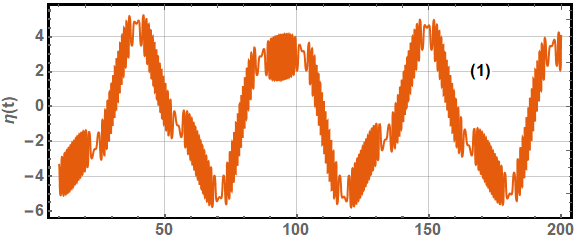} & %
\includegraphics[scale=0.38]{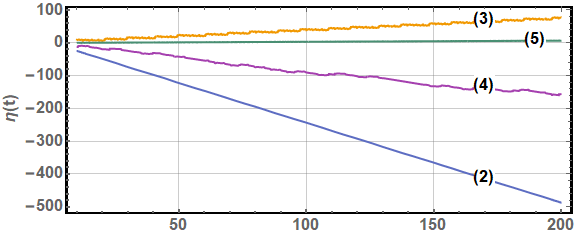}%
\end{tabular}
\caption{\label{figSP}\small The typical behaviour of the phases $\Psi(t)$, $%
\Phi (t) $, $\Theta(t)/t$, and $\protect\eta(t)$ in the zero-noise
single-mode model. The momentum $q_0$ is chosen such that $\protect\phi%
_{q_0}=\protect\pi/4$, the initial phase of $\Phi(t)$ is $\tilde{\Phi}_0=0.1$%
, the other parameters are different for each case: the case (1) $\Omega
=0.1/cs$, $\Omega X x_0=5$, the initial phase of $\Psi(t)$ is $\tilde{\Psi}%
_0=0.1$; (2) $\Omega =2.1/cs$, $\Omega X x_0=2.5$, $\tilde{\Psi}_0=\protect%
\pi/2+0.1$; (3) $\Omega =3.1/cs$, $\Omega X x_0=5$, $\tilde{\Psi}_0=\protect%
\pi-0.4$; (4) $\Omega =0.5/cs$, $\Omega X x_0=1$, $\tilde{\Psi}_0=-0.5$; (5) 
$\Omega =5.9$, $\Omega X x_0=4$, $\tilde{\Psi}_0=-\protect\pi/2-0.5$.}
\end{figure*}
In the trigonometric form of the equations of motion, one can clearly see the dynamic formation of the infinite (impenetrable) potential barrier for the polariton phase $\Psi(t)$ and, consequently, for the polariton amplitude $\cos^2\Psi(t)$ due to presence of the $\cot 2\Psi (t)$ function in eq.~(\ref{dotPhi}). This function forms a number of infinite barriers, which separate the phase space onto a set of permitted regions: $\frac{k\pi }{2}<\Psi (t)<\frac{(k+1)\pi }{2}$ with $k=0,\pm 1,\dots $ When $\Psi (t)$ comes close to the border of the permitted region, the cotangent function in eq.~(\ref{dotPhi}) becomes significant. Then the rapidly growing phase $\Phi (t)$ changes the sign of the derivative $\dot{\Psi}(t)$ in eq.~(\ref{dotPsi}).  Therefore presence of the nonlinear Hartree term in the equations of motions causes the splitting of the phase-space. 

The polariton wavefunction symmetry with respect to the phase shift $\Psi(t)\to\Psi(t)\pm \frac{\pi}{2} k$ ($k=1,2,\dots$) trivially follows from the same type of symmetry of the amplitude $\cos^2\Psi(t)$. Physically the role of the barrier can be described as follows. The repulsive nature of the barrier forbids the amplitude to become zero. In other words, the barrier separates the mixed polariton states from the pure states, i.e. the probability to find the system, when only the upper or the lower polariton is excited, equals to zero. This property disappears when the Hartree term is absent or when the vibration amplitude $x_0$ equals zero.

To demonstrate the repulsive properties of the barrier we solved eqs.~(\ref{dotPsi}) --~(\ref{doteta}) numerically for the model function $x(t)=\Omega Xx_{0}\cos \Omega t$, corresponding to a coherent excitation of a single vibration mode as, for example, in the coherent Raman spectroscopy experimental method~\cite{marowsky1992coherent}. In other words, we assume that only a single vibration degree of freedom with the parameters $\Omega $ and $X$ is excited with some amplitude $x_{0}$, and both the polariton subsystem and all other lower-frequency vibration modes (thermal bath) are neglected. The typical solutions of the eqs.~(\ref{dotPsi}) --~(\ref{doteta}) under the above choice of $x(t)$ are presented in fig.~\ref{figSP}. As expected the phase space of $\Psi$ is split onto permitted regions and the trajectory cannot cross the boundaries between the regions. For completeness we also calculated behaviour of the phases $\Phi(t)$, $\Theta(t)$, $\eta(t)$. Since the largest increment of the phase $\Phi (t)$ is defined by the cotangent, the linear asymptotic behaviour of $\Phi (t)$ is determined by these increments, which are regular in time due to quasiperiodicity of $\Psi (t)$. The solution for the small vibration frequency case (the case (1) in fig.~\ref{figSP} with $\Omega =0.1/cs$) is less typical. Here the short-scale dynamics is defined by the Rabi frequency $1/cs$ (the curves marked by (1) in fig.~\ref{figSP}), this dynamics is modulated by the oscillations of~$x(t)$.

Let us switch to another regime when one optically active vibration is resonantly activated by polariton, i.e. when the Rabi splitting equals to the vibration frequency.  In particular, we assume that the frequency $\Omega$ of some optically active vibration mode is in resonance with the Rabi splitting, i.e. $\Omega=\Lambda_+-\Lambda_-=1/cs$. The equations of motion for these optically active vibration modes without the bath terms read (see eq.~\ref{sigmadot}) 
\begin{equation}\label{sigmasimple}  
\dot{\sigma}_{m} =-\i \Omega \sigma_{m}+2\i\frac{\Omega X}{
N}\abs{ sC_1-cC_0}^2.
\end{equation}
At the initial time, before the formation of the dynamical barrier ($\sigma_{m}(0)=0$), the polariton wavefunction coefficients oscillate with the eigen polariton frequencies $\Lambda_{\pm}$. From eq.~(\ref{sigmasimple}) we can independently calculate the linearly growing with time resonant term, it is
\begin{equation}
\sigma(t)\propto\i t \frac{\Omega X}{
 N}e^{-\i \Omega t}s c \cos\Psi(0)\sin\Psi(0),
\end{equation} 
so that $\alpha(0;t)$ in eq.~(\ref{xoft}), which is proportional to the real part of $\sigma(t)$, can reach significant values. This linear behaviour at relatively large times is suppressed by the factors in eqs.~(\ref{C1noiseless}),~(\ref{C0noiseless}) proportional to $x(t)$, which eventually lead to increase of the detuning and thus constrain the resonant growth. Therefore when the Rabi frequency is in resonant with the frequency of the vibration the model predicts dynamical formation of the barrier, which separates the mixed polariton state from the pure states through the resonant growth of the amplitude at the infinite barrier. Note also that influence of any decoherence, noise processes and delocalization of the polariton packet in the regime of the vibronic level resonance pumping is inessential.

\section{Discussion and conclusion}

\label{Sec3} In the article, we derive the set of equations of vibration-assisted polariton motion (eqs.~\ref{Cdot1},~\ref{sigmadot}). To derive them we start from the conventional quadratic polariton Hamiltonian (eq.~\ref{HamP}) and the Holstein-like vibration Hamiltonian (eq.~\ref{Hvib}). The derivation method is based on the Dirac-Frenkel variation principle applied to the time-dependent basis of polariton states and on the multiconfiguration Hartree approach~\cite{WG2019,MMC1992}. In the section~\ref{sec3} we derive the set of equations of polariton motion (eqs.~\ref{sigmadot},~\ref{Cdot1}) and prove the correctness of the derivation procedure on the physical level of rigour. Note that the set of obtained equations for the given Hamiltonian $\hat{\mathcal{H}}$ (eq.~\ref{Hamtot}) is in exact one-to-one correspondence with the original Schr\"odinger equation, no approximations were made at the derivation. Obviously, the theory becomes approximate when one goes beyond the standard harmonic oscillator approximation for the vibrations and/or includes the non-resonant light-matter interaction terms into the polariton part.

To investigate the influence of the Hartree term we considered two particular examples. In section~\ref{Sec32} we estimate the polariton luminescence spectrum in the regime of large Rabi splitting and use the quasi-diagonal approximation~\cite{Fainberg22JPCA}). This allows us to linearise the equations of motion. Calculation of the polariton frequencies and the corresponding Hopfield coefficients is usually performed for the electronic Hamiltonian~ \cite{Hopfield58,Knoester_Mukamel89}. In this case, however, the dispersion equation for the polaritons cannot be reduced to the equation for the transverse eigenmodes of the medium~\cite{Fainberg18Advances,Fainberg19JPCC,Hau01}. To resolve this problem, the averaging of the Hopfield coefficients with respect to the low frequency optically active vibrations has been done in Ref.~\cite{Fainberg22JPCA}, which made it possible to get the dispersion equation coinciding with the equation for the transverse eigenmodes. In present approach this procedure can be performed consistently as we demonstrated in the section~\ref{Sec32}.  The resulting spectra accurately reproduce the physical properties of the spectra, see fig.~\ref{FigDiagr}~b. Note that the theory catches the effect of the Stokes shift in the polariton luminescence spectra. Namely, in the region of large $q$, where the polariton particle has a large exciton weight the position of the luminescence maximum is red-shifted by the factor $\omega_{st}$ from the position of the polariton energy $\Lambda_{-q}$. In the opposite regime, when the polariton particle is essentially a photon, the fluorescence peak coincides with the energy of the photon component of the polariton. Note also that the energy shift can be smaller than the maximal possible shift equal to $\omega_{st}$. The shift value depends on the particular distribution of the polariton wavefunction in $q$-space.  The effect of molecular Stokes shift on polariton spectra for strong coupling was seen and discussed in a number of experimental works~\cite{LBVAS1999,TBKPN2017,HPTTGT2021}. Our theory can serve as a basis for the heuristic model formulated in Ref.~\cite{TBKPN2017} qualitatively explaining why the Stokes shift in resonant cavities filled with R6G:PMMA is larger than that in the same dye-doped films deposited on glass. Moreover, our theory explains also the luminescence spectrum narrowing of the R6G:PMMA film in the cavity with respect to the luminescence spectrum of the same film deposited on glass~\cite{TBKPN2017}.

Our theoretical approach originates from the theory of diabatic and adiabatic processes in quantum systems with the avoided-energy level crossing~\cite{K1992,WG2019,MMC1992}. In such theories, the temporal switching between the energy branches is usually introduced through a time-dependent external parameter~\cite{K1992}. According to our consideration, in the polariton-vibration system, the time-dependence comes from the vibrations dynamically activated by the polariton mean-field molecular. Such kind of equations are typical for the many-particle theories~\cite{WG2019,MMC1992}. In section~\ref{sec21} we considered the polariton wavefunction behaviour in the vicinity of AP. In particular, we showed that when an optically-active vibration is resonantly excited, the polariton particle in the AP vicinity exists only in a mixed quantum state of the upper and lower polariton. This state is prevented from further decay into a pure upper or lower polariton state by the infinite energy barriers of a dynamical origin.

In the article, we provide a rigorous derivation of the equations governing the vibration-assisted evolution of the polariton wavefunction. There are a number of special problems, which were left aside from the focus of this article, such as revealing the vibronic progression in the polariton spectra, polariton diffusion along the dispersion curve, and the Bose-Einstein condensation description. In a more general context, our theory can be useful in the construction of rigorous approaches for the purposes of multidimensional spectroscopy~\cite{FMSWMBPZ2021}. Though as noted, our theory describes the effects of molecular Stokes shift in polariton spectra at strong coupling~\cite{TBKPN2017}, the same effects can be described within our previous approach~\cite{Fainberg22JPCA}. However, the fundamental advantage of the developer here theory is in the rigorous formulation of nonlinear equations of motion. Therefore, its significance goes far beyond the explanation of the spectral Stokes shift at the strong light-matter coupling. We expect that the theory will lead us to new, associated with the nonlinearity, manifestations of the collective behaviour in polaritonic and similar systems, for example, mutual synchronization of interacting oscillators, oscillation death etc.~\cite{Pikovskyf01}. These issues will be considered elsewhere.

\begin{acknowledgments}
The work was supported by the Ministry of Science \& Technology of Israel (grant No. 79518) and the grant RA1900000633 for cooperation between the Ariel University and the Holon Institute of Technology. Authors thank Hashem Zoubi for useful discussions. V.O. thanks Boris Gutkin for providing additional financial support. Authors thank Eugene Kanzieper and Roman Riser for providing access to the high-performance computational resources during the initial stage of the research.
\end{acknowledgments}

\appendix

\section{Derivation of the Polariton Fluorescent signal}

\label{AppB} According to the optic version of the Wiener-Khintchine
theorem, the frequency-resolved signal, $S(\omega )$, is expressed in terms
of the radiation field autocorrelation function, 
 \begin{multline}
S(\omega )=\frac{1}{\pi }\mathrm{Re}\int_{-\infty }^{+\infty }d\tau
\int_{-\infty }^{\tau }d\tau ^{\prime }e^{-\i \omega (\tau -\tau
')}\\\times \left\langle E^{(-)}(\bm r,\tau
)E^{(+)}(\bm r,\tau ^{\prime })\right\rangle .
\end{multline}
The correlation function is the quantum expectation value ($\tau ^{\prime
}<\tau $), 
\begin{multline}
\left\langle E^{(-)}(\bm r,\tau )E^{(+)}(\bm r,\tau')\right\rangle
\\ =\text{Tr}\;\left[ E^{(-)}(\bm r,\tau )E^{(+)}(\bm r,\tau ^{\prime })\rho
_{Tot}(\tau |\tau ^{\prime })\right] ,  \label{EEcorr}
\end{multline}
where $\rho _{Tot}$ is the density matrix of the total system, including the
external field. The quantized outer electric field, $E^{(\pm )}(\tilde{q},t)
$ defined in the space of wavevectors $\tilde{q}$, which tangential component equals $q_{||}$, is 
\begin{multline}
E^{(-)}(\tilde{q},t)+E^{(+)}(\tilde{q},t)\\=-\i \int d\omega \frac{\sqrt{%
\hbar \omega }}{c}\left[ F_{\tilde{q}}^{\dag }(\omega )e^{\i \omega
t}-F_{\tilde{q}}(\omega )e^{-\i \omega t}\right] ,
\end{multline}%
where $F_{\tilde{q}}(\omega )$ is the field operator, which also includes the dispersion relation $\delta (\omega-\omega _{\tilde{q}})$ in the outer space. The Hamiltonian term of the outer field
is $H_{out}=\sum_{\tilde{q}}\int d\omega \hbar \omega F_{\tilde{q}}^{\dag }(\omega)F_{\tilde{q}}(\omega )$. The coupling $V$ between the outer and the cavity light
modes we describe in the minimal coupling assumption, i.e. 
\begin{equation}\label{VI}
V=\i \frac{\hbar \kappa }{2\pi }\sum_{q}\int d\omega \left[
A_{q}F_{\tilde{q}}^{\dag }(\omega )-A_{q}^{\dag }F_{\tilde{q}}(\omega )\right] ,
\end{equation}%
where the photon leakage coefficient, $\kappa $, is assumed to be $q$%
-independent. The tangential component $q_{||}$ of the wavevectors $\tilde{q}$ and $q$ coincide, while $z$-component of $q$ is fixed by the photon mode excited in the cavity (fig.~\ref{fig2a} a) and $z$-component of $\tilde{q}$ is determined by the emitted photon energy and by $q_{||}$. Thus the sum in eq.~(\ref{VI}) effectively runs over $q_{||}$.

The density matrix evolution satisfies the von Neumann's equation $\dot{\rho}%
_{Tot}=-\frac{\i}{\hbar}[H_{out}+V+\hat{\mathcal{H}}+H_{excit},\rho_{Tot}]$,
which in the interaction picture transforms into the equation $\dot{\rho}%
_{Tot,I}=-\frac{\i}{\hbar}[V_I,\rho_{Tot,I}]$ with 
\begin{multline}
V_I(t|t_0)=\i \frac{\hbar\kappa}{2\pi} \sum_q \int d\omega  \Big[A_q(t|t_0)
F_{\tilde{q}}^\dag(\omega)e^{-\i\omega (t-t_0)}\\-A^\dag_q(t|t_0) F_{\tilde{q}}(\omega)
e^{\i\omega (t-t_0)}\Big],
\end{multline}
and the time-dependent photon operators, 
\begin{multline}\label{Aqt}
A_q(t|t_0)=Te^{\i\int_{t_0}^t(\hat{\mathcal{H}}(\tau)+H_{exit}(\tau))d\tau/
\hbar}\\\times A_{q}\;Te^{-\i\int_{t_0}^t(\hat{\mathcal{H}}(\tau)+H_{exit}(\tau))d\tau/
\hbar}.
\end{multline}  
The $H_{excit}$ part of the total Hamiltonian is responsible for creation of
polaritons by the external classical pump field.

The density matrix expanded up to the second order over the interaction $V_I$
is 
\begin{widetext}
\begin{equation}
\rho_{Tot,I}(t|t_0)=\rho_{Tot,I}^{(0)}-\frac{i}{\hbar}\int_{t_0}^t d\tau
[V_I(\tau|t_0),\rho_{Tot,I}^{(0)}]+\frac{1}{\hbar^2}\int_{t_0}^t
d\tau_1\int_{t_0}^{\tau_1} d\tau_2
[V_I(\tau_1|t_0),[V_I(\tau_2|t_0),\rho_{Tot,I}^{(0)}]]
\end{equation}
\end{widetext}
This expansion has to be substitute under the trace in eq.~(\ref{EEcorr}).
Since the product $E^{(-)}(\bm r,\tau)E^{(+)}(\bm r,\tau^{\prime })$ already
contains the $F_q^\dag F_q$ and we assume that initially the outer light
modes are empty, we obtain that $\text{Tr}\;[F_q^\dag
F_q\rho_{Tot,I}^{(0)}]=0$. The non-trivial combinations of operators has to
contain the traces $\text{Tr}\;[F_q^\dag F_q F_q^\dag
\rho_{Tot,I}^{(0)}F_q]=1$. These combinations are generated by the products $%
F_q^\dag F_q V_I^{(-)}(\tau_1|t_0)\rho_{Tot,I}^{(0)} V_I^{(+)}(\tau_2|t_0)$
and $F_q^\dag F_q V_I^{(-)}(\tau_2|t_0)\rho_{Tot,I}^{(0)}
V_I^{(+)}(\tau_1|t_0)$. The first contribution to the signal amplitude, $%
S^{1}(\omega)$ measured at the position marked by the radius vector $\bm r$,
has the form 
\begin{widetext}
\begin{multline}
S^{1}(\omega)=\frac{1}{2\pi\hbar}\mathrm{Re}\int_{-\infty}^{+\infty}d\tau
\int_{-\infty}^{\tau}d\tau^{\prime }\int d\omega_1 \sqrt{\omega_1}\int
d\omega_2 e^{\i\omega_1 \tau}\sqrt{\omega_2}e^{-\i\omega_2 \tau^{\prime
}}e^{-\i\omega(\tau-\tau^{\prime })} \\
\times\int_{\tau^{\prime }}^\tau d\tau_1\int_{\tau^{\prime }}^{\tau_1}
d\tau_2 \sum_{\tilde{q},\tilde{q}'}e^{\i(\tilde{\bm q} -\tilde{\bm q}^{\prime })\bm r}\Big[ \text{Tr%
}\;\;F_{\tilde{q}}^\dag(\omega_1) F_{{\tilde{q}}^{\prime }}(\omega_2)
V_I^{(-)}(\tau_1|\tau^{\prime })\rho_{Tot,I}^{(0)}
V_I^{(+)}(\tau_2|\tau^{\prime }) \\
+\text{Tr}\;\;F_{\tilde{q}}^\dag(\omega_1) F_{{\tilde{q}}^{\prime }}(\omega_2)
V_I^{(-)}(\tau_2|\tau^{\prime })\rho_{Tot,I}^{(0)}
V_I^{(+)}(\tau_1|\tau^{\prime })\Big]
\end{multline}
\end{widetext}
Since the detector is positioned far from the sample, the major contribution
to the signal comes from the terms with ${\tilde{q}}={\tilde{q}}^{\prime }$, thus the signal
amplitude measured in the direction marked by a given vector ${\tilde{q}}$ is determined by the tangential component $q_{||}$
\begin{multline}
S^{1}_{q_{||}}(\omega)=\frac{\hbar\omega_{\tilde{q}} \kappa^2}{2\pi^2} \mathrm{Re}
\int_{-\infty}^{+\infty}d\tau \int^{\tau}_{-\infty}d\tau^{\prime
-\i(\omega-\omega_{\tilde{q}})(\tau-\tau^{\prime })} \\
\times\mathrm{Re}\int_{\tau^{\prime }}^\tau d\tau_1\int_{\tau^{\prime
}}^{\tau_1} d\tau_2 \text{Tr}\;\;F_{\tilde{q}}^\dag(\omega_{\tilde{q}}) F_{\tilde{q}}(\omega_{\tilde{q}})
A_q^\dag(\tau_1|\tau^{\prime })\\\times F_{\tilde{q}}^\dag(\omega_{\tilde{q}}) \rho_{Tot,I}^{(0)}
A_{q}(\tau_2|\tau^{\prime })F_{\tilde{q}}(\omega_{\tilde{q}})e^{-\i\omega_{\tilde{q}} (\tau_1-\tau_2)}.
\end{multline}
After taking the trace and summing up all terms we obtain the expression 
\begin{multline}  \label{signalqomega}
S_{q_{||}}(\omega)=\frac{\hbar\omega_{\tilde{q}} \kappa^2}{\pi^2} \mathrm{Re}
\int_{-\infty}^{+\infty}d\tau \int_{-\infty}^{\tau}d\tau^{\prime
-\i(\omega-\omega_{\tilde{q}})(\tau-\tau^{\prime })} \\
\times\mathrm{Re}\int_{\tau^{\prime }}^\tau d\tau_1\int_{\tau^{\prime
}}^{\tau_1} d\tau_2 e^{-\i\omega_{\tilde{q}}(\tau_1-\tau_2)}\\\times \bra{q,\tau'} A_q^\dag(\tau_1|\tau^{\prime })
A_q(\tau_2|\tau^{\prime })\ket{q,\tau'}.
\end{multline}
The quantum state $\ket{q,\tau'}$ here is a state with a previously (at
instant of time $\tau^{\prime }$) created polariton, which quantum amplitude
has a non-zero projection onto the state with the wavevector $q$ during the
time of the polaritonic system evolution. Note, that the creation of
polariton is governed by the $H_{excit}$ Hamiltonian. Its structure $%
H_{excit}=\hbar\sum_q \left(r_{ex}(q,t) P_q^\dag+r^{(+)}_{ex}(q,t)
Q_q^\dag\right)+c.c.$ describes creation of the upper and the lower $q$%
-polaritons with some time-dependent rates $r_{ex}$ and $r^{(+)}_{ex}$,
respectively. 
To have a non-zero quantum average $\bra{q,\tau'} A_q^\dag(\tau_1|\tau^{%
\prime }) A_q(\tau_2|\tau^{\prime })\ket{q,\tau'}$ the polartiton has to be
excited twice. The form of the excitation term depends on the particular
realization of the excitation mechanism. Assume that the excitation happen
for the lower polaritons only ($r^{(+)}_{ex}\equiv 0$). Thus, explicitly,
after expansion over $H_{exit}$ we write 
\begin{multline}  \label{factorization}
\bra{q,\tau'} A^\dag_q(\tau_1|\tau^{\prime }) A_q(\tau_2|\tau^{\prime })%
\ket{q,\tau'}= \bar{r}_{ex}^2 \\\times \sum_{q^{\prime }}\int_{\tau^{\prime
}}^{\tau_1}\frac{r_{ex}(q^{\prime },\tau_1^{\prime })}{\bar{r}_{ex}}%
d\tau_1^{\prime }\cdot \bra{0} P_{q^{\prime }} Te^{\i\int_{\tau_1^{\prime
}}^{\tau_1}\hat{\mathcal{H}}(\tau)d\tau/\hbar}A^\dag_q\ket{0} \\
\times \sum_{q^{\prime \prime }}\int_{\tau^{\prime }}^{\tau_2}\frac{%
r_{ex}(q^{\prime \prime },\tau_2^{\prime })}{\bar{r}_{ex}} d\tau_2^{\prime
}\cdot \bra{0}A_q Te^{-\i\int_{\tau_2^{\prime }}^{\tau_2}\hat{\mathcal{H}}%
(\tau)d\tau/\hbar}P_{q^{\prime \prime }}^\dag\ket{0}.
\end{multline}
Here we introduced a typical excitation rate constant $\bar{r}_{ex}$. Such
factorization (eq.~\ref{factorization}) allows us further simplification of the expression for the signal (eq.~\ref{signalqomega}) by taking the Fourier transform of each component, 
\begin{multline}\label{G1AppB}
G_1(\omega_2,\omega_3)= \iint_{-\infty}^{+\infty}d\tau_1 d\tau^{\prime
\i\omega_2\tau_1+\i\omega_3\tau^{\prime }}\sum_{q'}\int_{\tau'}^{\tau_1}d\tau_1' \\\times  \frac{r_{ex}(q^{\prime },\tau_1^{\prime })}{%
\bar{r}_{ex}}\cdot \bra{0} P_{q^{\prime }}
Te^{\i\int_{\tau_1^{\prime }}^{\tau_1}\hat{\mathcal{H}}(\tau)d\tau/\hbar}A^%
\dag_q\ket{0},
\end{multline}  
\begin{multline}  \label{G2AppB}
G_2(\omega_4,\omega_5)= \iint_{-\infty}^{+\infty}d\tau_2 d\tau^{\prime
\i\omega_4\tau_2+\i\omega_5\tau^{\prime }}\sum_{q^{\prime \prime
}}\int_{\tau^{\prime }}^{\tau_2}d\tau_2'\\\times \frac{r_{ex}(q^{\prime \prime
},\tau_2')}{\bar{r}_{ex}}\cdot \bra{0}A_q
Te^{-\i\int_{\tau_2^{\prime }}^{\tau_2}\hat{\mathcal{H}}(\tau)d\tau/%
\hbar}P_{q^{\prime \prime }}^\dag\ket{0},
\end{multline}
such that the overall expression for the signal becomes 
\begin{multline}  \label{SqAppB}
S_{q_{||}}(\omega)= \hbar\omega \frac{\bar{r}_{ex}^2\kappa^2%
}{\pi } \delta(\omega -\omega_{q_{||}})\mathrm{Re} \int_{-\infty}^{+\infty} d\omega_3 \\
\times \Big( %
G_1(-\omega,\omega_3)G_2(\omega,-\omega_3) +
G_1^*(\omega,\omega_3)G_2^*(-\omega, -\omega_3) \Big).
\end{multline}
At the derivation of the last expression we implied strict conservation of
energy and committed the off-resonant contributions.

\section{Estimation of the Hartree term in eq.~(\ref{sigmadot})}
\label{AppD}
The Hartree term in eq.~(\ref{sigmadot}), $\frac{\Omega_\mu X_\mu}{N}\chi_m(t|\bm\sigma)$, can be estimated from above  by means of the Cauchy–Bunyakovsky–Schwarz inequality (the square of a sum is less or equal to the sum of squares). Namely,  
{\small\begin{multline}
\chi_m(t|\bm\sigma)\stackrel{\text{(i)}}=\;\abs{\sum_q e^{\i q m}\bigg(\sin\phi_q\mathcal{C}
(\bm\sigma ,q,1|t) -\cos\phi_q \mathcal{C}
(\bm\sigma ,q,0|t)\bigg)}^2
\\\stackrel{\text{(ii)}}{\lesssim}\;
\abs{\sum_{q}e^{\i q m}\sin\phi _{q}\mathcal{C}
(\bm\sigma ,q,1|t)}^2 +\abs{\sum_{q}e^{\i q m}\cos\phi _{q}\mathcal{C}(\bm\sigma ,q,0|t)}^2\\\stackrel{\text{(iii)}}{\le}\;
N\sum_{q}\bigg( \sin ^{2}\phi _{q}\left\vert \mathcal{C}(\bm\sigma ,q,1|t)\right\vert ^{2}+\cos ^{2}\phi _{q}\left\vert \mathcal{C}(\bm\sigma ,q,0|t)\right\vert ^{2}\bigg),
\end{multline}}
the equivalence (i) follows directly from the definitions of the Hartree force (eqs.~\ref{chi0},~\ref{rho},~\ref{vmatrix}), the approximate inequality (ii) is achieved after omitting the highly oscillating terms, while the inequality (iii) represents the Cauchy–Bunyakovsky–Schwarz inequality.  The latter inequality becomes close to the exact equivalence for a narrow Gaussian distribution of $\mathcal{C}
(\bm\sigma ,q,u|t)$ in the $q$-space. The latter argument allows us to use the exact upper border as the estimator of the Hartree term $\chi_m$, see section~\ref{Sec321}. Note also that the diagonal approximation, i.e. when the exponential function $e^{\i (q-q')m}$ is replaced by unity, ends up in the same resulting expression.

\section{Estimation of the transition rate in the balance equations}

\label{AppC} The sum of the equation of motion (eq.~\ref{Cdot1}) taken with the multiplier $\mathcal{C}(\bm \sigma,q,u|t)$ with its complex conjugation yields the equation for the density matrix $\rho(q,u|q',u';t)=\mathcal{C}^* (\bm \sigma,q',u'|t) \mathcal{C} (\bm \sigma,q,u|t)$ of the polariton subsystem (we took into account that $\mathcal{N}(t)=1$). The corresponding von Neumann's equation reads 
\begin{widetext}
\begin{equation}  \label{vonNeumann1}
\frac{\partial}{\partial t}\rho(q^{\prime },u^{\prime }|q,u;t)
=-\i\sum_{q^{\prime \prime },u^{\prime \prime }}\Bigg[h(q^{\prime
},u^{\prime }|q^{\prime \prime },u^{\prime \prime };t)\rho(q^{\prime \prime
},u^{\prime \prime }|q,u;t)-\rho(q^{\prime },u^{\prime }|q^{\prime \prime
},u^{\prime \prime };t)h(q^{\prime \prime },u^{\prime \prime }|q,u;t)\Bigg].
\end{equation}
\end{widetext}
with the hermitian matrix $h(q^{\prime },u^{\prime }|q^{\prime \prime
},u^{\prime \prime };t)$, 
{\small \begin{eqnarray}
h(q^{\prime },u'|q,u;t)&\equiv&h_0(q^{\prime },u^{\prime
}|q,u)-h_1(q^{\prime },u'|q,u;t); \\
h_0(q^{\prime },u^{\prime }|q,u)&\equiv&\left((1-u^{\prime
})\Lambda_{-q^{\prime }}+u^{\prime }\Lambda_{+q^{\prime
}}\right)\delta_{q^{\prime },q}\delta_{u^{\prime },u}; \\
h_1(q^{\prime },u^{\prime }|q,u;t)&\equiv&\alpha(q'-q )\big[\bm %
v_{q^{\prime },q}\big]_{u^{\prime },u}.
\end{eqnarray}}
The hermitisity of $h(q^{\prime },u^{\prime }|q^{\prime \prime },u^{\prime
\prime };t)$ follows from the symmetries $\alpha^*(q-q^{\prime \prime
})=\alpha(q^{\prime \prime }-q )$ and $\big[\bm v_{q,q^{\prime }}\big]%
_{u,u^{\prime }}=\big[\bm v_{q^{\prime },q}\big]_{u^{\prime },u}$.

Our aim is to obtain a system of equations for the diagonal entries of the
density matrix. To this end we use the method proposed by Zwanzig~\cite{Z1964} and represent the density operator as a sum of
diagonal and off-diagonal terms by means of the projection operator $\hat{D}$
(a three-dimensional tensor with the entries $\hat{D}_{m,n,k}=\delta_{m,n}%
\delta_{n,k}$ and satisfying the properties $\hat{D}^2=\hat{D}$, $(\mathds{1}%
-\hat{D})^2=(\mathds{1}-\hat{D})$), such that $\rho =\hat{D}\rho+(\mathds{1}-%
\hat{D})\rho$. Using the obvious properties $\hat{D}[h,\hat{D}\rho]=[\hat{D}%
h,\hat{D}\rho]=0$, which holds for any $h$ and $\rho$, we obtain 
\begin{eqnarray}  \label{dotrho1}
\hat{D}\dot{\rho}&=&-\i \hat{D}[h,(\mathds{1}-\hat{D})\rho]; \\
(\mathds{1}-\hat{D})\dot{\rho}&=&-\i(\mathds{1}-\hat{D})[h,\hat{D}\rho]\nonumber\\&&-\i (
\mathds{1}-\hat{D})[h,(\mathds{1}-\hat{D})\rho].
\end{eqnarray}
It is natural to assume that the initial density matrix has the diagonal
entries only, so that $(\mathds{1}-\hat{D})\rho(0)=0$. Therefore the
solution for $(\mathds{1}-\hat{D})\rho(t)$ is given by the integral 
\begin{equation}  \label{rho2Sol}
(\mathds{1}-\hat{D})\rho(t)=-\i (\mathds{1}-\hat{D})\int_0^t
u^\dag(\tau,t)[h(\tau),\hat{D}\rho(\tau)]u(\tau,t)d\tau,
\end{equation}
with the unitary matrix 
\begin{equation}
u(\tau,t)=T\exp\left[-\i\int_t^\tau h(\tau_1)d\tau_1\right]
\end{equation}
Substitution of the expression eq.~(\ref{rho2Sol}) into eq.~(\ref{dotrho1})
gives rise to the balance equations for the diagonal entries of the density
matrix in the form 
\begin{multline}\label{diagvonNeumann}
\hat{D}\dot{\rho}=-\hat{D}\int_0^t d\tau \bigg[(\mathds{1}-\hat{D}) h(t),\\
u^\dag(\tau,t)\big[(\mathds{1}-\hat{D})h(\tau),\hat{D}\rho(\tau)\big]u(\tau,t)\bigg].
\end{multline}  

Having formulated the equation for the diagonal entries of the density matrix (eq.~\ref{diagvonNeumann}) we have to make the thermodynamic averaging. To perform the averaging, we can assume that the vibration degrees of freedom equilibrate very fast between the optical transitions. This allows us, first, to make the factorization of the expectation value $<h(q^{\prime
},u^{\prime }|q^{\prime \prime },u^{\prime \prime };t)\rho(q^{\prime \prime
},u^{\prime \prime }|q,u;t)>=<h(q^{\prime },u^{\prime }|q^{\prime \prime
},u^{\prime \prime };t)><\rho(q^{\prime \prime },u^{\prime \prime }|q,u;t)>$; second, to draw the density matrix out the time integration; and to use the limiting distribution for $\alpha_m$, as in section~\ref{Sec32} (eq.~\ref{Probmeasure}), which is the Gaussian distribution with some mean value $\bar{\alpha}_m$, 
{\small \begin{equation}\label{palpha}
P(\alpha_m)=\frac{1}{\sqrt{2\pi  Nk_N \omega_{st}  k_BT}}\exp\left[-\frac{
(\alpha_m- \bar{\alpha}_m)^2}{2 Nk_N\omega_{st} k_BT}\right].
\end{equation}}
From the definition of $<\alpha(0)>$, which is $<\alpha(0)>=N^{-1}\sum_m \bar{\alpha}_m$, we can also approximately replace each $\bar{\alpha}_m$ by $<\alpha(0)>$. One can show, that under above assumption the balance equation (eq.~\ref{diagvonNeumann}) for the lower polariton branch, eventually, reduces to the form  
\begin{multline}\label{balanceApp}
\frac{d}{dt}<\rho(q,0|q,0;t)>=\sum_{q'}\mathcal{K}(q,q')\\\times \bigg( <\rho(q,0|q,0;t)>-<\rho(q',0|q',0;t)>\bigg),
\end{multline}  
with the time-independent kernel $\mathcal{K}(q,q')$, which is approximately calculated as 
\begin{multline}\label{kernel}
\mathcal{K}(q,q')\approx \Big<\int_0^t d\tau\left[(\mathds{1}-\hat{D}) h(t)\right]_{q,q^{\prime
}}\left[(\mathds{1}-\hat{D})h(\tau) \right]_{q^{\prime },q}\\ \times
\left(u_{q^{\prime },q^{\prime }}^\dag(\tau,t) u_{q,q}(\tau,t) +u_{q^{\prime
},q^{\prime }}(\tau,t) u_{q,q}^\dag(\tau,t) \right)\Big> .
\end{multline}
The rough estimation of the transition rate can be done by averaging of the leading term $\frac{1}{N^2}\sum_m\alpha_m^2$ in the kernel eq.~(\ref{kernel}) with respect to the probability measure eq.~(\ref{palpha}).  It gives the estimation $\abs{\mathcal{K}(q,q')}\propto \sqrt{k_N \omega_{st}  k_BT/N }$.

\end{document}